\documentclass[a4paper,10pt]{article}       

\usepackage{lscape}

\usepackage{subcaption}

\usepackage[utf8]{inputenc}
\usepackage{amsmath,amssymb}

\usepackage{graphicx}
\usepackage[space]{grffile}

\usepackage{array}

\usepackage{algorithm}
\usepackage{algpseudocode}

\usepackage{epstopdf}

\usepackage{pifont}

\begin{document}

\title{On the Complex Network Structure of Musical Pieces: Analysis of Some Use Cases from Different Music Genres}

\author{Stefano Ferretti\\
	  Department of Computer Science and Engineering, University of Bologna\\
	  Mura Anteo Zamboni 7\\
	  Bologna, Italy\\
	  s.ferretti@unibo.it
}

\date{}
\maketitle

\begin{abstract}
This paper focuses on the modeling of musical melodies as networks. Notes of a melody can be treated as nodes of a network. Connections are created whenever notes are played in sequence. We analyze some main tracks coming from different music genres, with melodies played using different musical instruments. We find out that the considered networks are, in general, scale free networks and exhibit the small world property. We measure the main metrics and assess whether these networks can be considered as formed by sub-communities. Outcomes confirm that peculiar features of the tracks can be extracted from this analysis methodology. This approach can have an impact in several multimedia applications such as music didactics, multimedia entertainment, and digital music generation.
\end{abstract}

\section{Introduction}

The recent advances in information retrieval techniques, big data analysis and complex network methodological tools foster novel approaches to the musical domains, ranging from music classification, categorization to automatic ge\-ne\-ration \cite{Angeler2016,icme16,Oord:2013,Patra:2013,thickstun2016learning,videocoding,videocoding2,6890565}.
It has been recognized that network science can be employed to represent music as a network \cite{Ferretti2017271,Liu2010126}.
This is a consequence of the motto ``everything as a network'', based on which, several types of real and digital systems are represented and studied as complex networks. Examples range from food webs, human language to communication and mobile networks \cite{Boccaletti2006175,Cancho2261,Cong2014598,Ferretti20131631,Ferretti2013481,Grabska,Newman:2010,pardo}.

In the musical domain, networks can be constructed to model melodies (and related harmonies), with nodes corresponding to musical notes and edges corresponding to their co-occurring connections \cite{Ferretti2017271,Liu2010126}.
This alternative view of a musical piece can give a graphical representation that provides a first sketch of how complex or simple is the melody itself. But there is much more: sophisticated analyses can be made to measure mathematical metrics which characterize the network. This allows obtaining important insights on the net and the corresponding musical pieces. Moreover, based on these metrics it is possible to compare different pieces, artists and music genres.

The approach differs from works related to audio analysis, where audio contents are manipulated so as to obtain measures related to similarity, statistical proportions of music attributes or other metrics enabling classification \cite{Berenzweig:2004,Manaris:2005}.
In fact, the network is created starting from a symbolic representation of the audio data, i.e., using a music score sheet. This clearly eases the ana\-lysis. 
It has been recognized that most musicological concepts such as melodic and harmonic structure are easier to investigate in the symbolic domain, and usually more successful \cite{Knopke2011}.
The reader can refer to \cite{Fu:2011} for a comprehensive review on audio and music based classification schemes presented in the literature.

Given a musical score sheet, we will focus on melodic lines, e.g.~the main melody of a classical music composition or a solo performed by a musician in a jazz tune.
The rationale is to exclude the repetitive parts of the track and concentrate to the main and variable parts of a music song.
In jazz and rock music, a solo is considered as one of the prominent parts of a musical piece, since it allows identifying the technical artistic skills of a performer. 
Indeed, it is a common claim that each jazz and rock musician has its own typical ``musical language'', composed of preferred ``licks'' (i.e.,~recurrent patterns and sequences of notes), scales, rhythmic patterns. 

The representation of a melodic line as a network allows identifying the main characteristics of the music style of an artist, since the obtained network harnesses the musical units (i.e.,~notes, chords, rests) and their relations. In fact, emergent properties due to the interactions between such music elements can be decoded and analyzed. 

As a proof of concept, we analyze musical solos and main melodies of famous compositions and improvisations of different composers and performers, playing different instruments. 
We measure the main properties of the correspondent networks and discuss on the main metric values measured through the net analysis, and their meaning from a music analysis perspective.
The considered tracks have been widely analyzed in musical terms; thus, the comparison between the general musical aspects and the measures from their network representation facilitates a musical interpretation of the obtained outcomes.

Results provide interesting insights. The structure of the network gives an idea of the complexity of the correspondent musical melody. It turns out that we are dealing with scale-free networks, meaning that they show a degree distribution that can be approximated with a power law function \cite{Ferretti2013481}. Scale free networks possess a high majority of nodes having small degrees; however a non negligible portion of nodes have higher degrees \cite{Newman:2010}. It is possible to identify hubs (i.e., nodes with high degrees) that have a high number of connections in the network. This means that musicians do have some preferred notes that are exploited during the composition of a melody track.
This approach easily permits also detecting which are those nodes that have a main influence on the connectivity of the network (i.e., nodes with a high betweenness value). These are notes one should pass through to go from one part of the network to another.

Other metrics, such as the diameter, average path length, net density, clustering coefficient, concur in the identification of the complexity of the melody. These values can be also employed to identify if the network is a small world, i.e., 
a net in which most nodes are not neighbors of one another, but the neighbors of any given node are likely to be neighbors of each other and a small number of hops is required to go from one node to another.
Moreover, the modularity of a network can be measured, that describes if the network is clearly formed by some main sub-networks (communities) which are densely connected with respect to the network. Such a feature would witness a preference of a musician in playing certain groups of notes together in a given part of the melody.
Finally, it is possible to identify those notes' pairs that are frequently played in sequence. This promotes the identification of important patterns that are widely employed by a given musician in its ``musical language''.

The presented approach can be exploited to discriminate among the main features of a musician, a music track or even a music genre.
Thus, it can be employed as a tool inside a plethora of multimedia applications concerned with music classification, categorization, automatic generation of digital music, didactic scenarios and multimedia entertainment \cite{Bell:2011,GranSteed:14,KellerSTME13,LichtenwalterLC10}.

The remainder of this paper is structured as follows. 
Section \ref{sec:musicnet} presents an approach to model musical pieces as networks.
Section \ref{sec:metrics} discusses on the main metrics of interest that characterize the networks build from the musical tracks.
Section \ref{sec:ex} provides and analysis on some practical examples coming from melodies of well-known musical tracks.
Section \ref{sec:eval} shows some aggregate results of metrics of interest, obtained from a large set of melodic lines related to musical solos of contemporary musicians.
Section \ref{sec:conc} concludes the paper with some final remarks.

\section{Modeling Musical Pieces as Networks}\label{sec:musicnet}

\subsection{From a music sheet to a network}

Starting from a music sheet, a corresponding network can be built as follows. Nodes of the network correspond to specific notes. The note can be a single one, a rest or a chord, i.e., a group of notes played simultaneously. Each node has a label associated to it.
Labels vary depending on the type of note.
In case of a single note, the related node has a label composed of the note pitch, octave and duration. A ``rest node'' is labeled with the duration of the rest. Finally, nodes corresponding to chords are labeled with the pitch, octave and duration of each note composing the chord.
Links are associated to nodes that correspond to notes played in sequence in the sheet.

\begin{figure}
\centering
  \includegraphics[width=.7\linewidth]{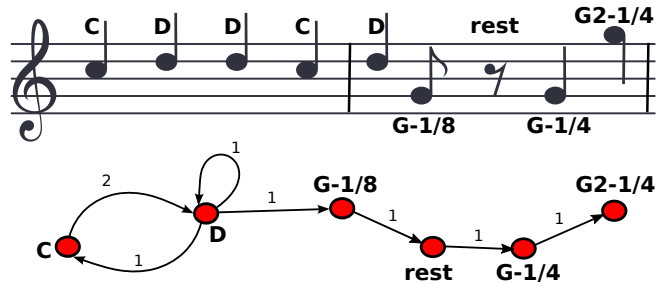}
\caption{Example of melodic line mapped to a network}
\label{fig:netExample}
\end{figure}

Let consider the example shown in Figure \ref{fig:netExample}, where a simple music score sheet of a melodic line is depicted together with the correspondent network. The text label reported over each note is the name of the note. Different notes are mapped into different nodes, that are labeled with the note name, as mentioned (notice that we adopt here a simplified notation, and omit duration and octave). Weights are assigned to links, counting the amount of occurrences of the corresponding notes pair. Thus, a link is created from the $C$ node to $D$, $(C, D)$, since the first note on the sheet is a $C$, followed by a $D$. Then, a self loop $(D, D)$ is added to the network, since the third note on the sheet is a $D$, again. The fourth note is a $C$, that corresponds to the $(D, C)$ link. 
A second occurrence of the $(C, D)$ pair increases the weight associated to that link.
Then, there is a sequence of links $(D, G-1/8), (G-1/8, \text{rest}), (\text{rest}, G-1/4), (G-1/4, G2-1/4)$. Note that there are three different nodes for the $G$ notes, since $G-1/8, G-1/4$ have the same pitch (i.e.,~$G$) but different duration (the first $G$ is a eighth note, while the second one is a quarter); moreover, $G2-1/4$ is an octave higher than other two $G$ notes.

As a final remark, we notice that nodes and links might be enriched with further information related to specific musical aspects, e.g., a ``legato'' sequence or better, the percentage of links that derive from legato notes. However, in this study we do not consider these additional features.

The additional information added to nodes, or links, allows also to reconstruct the original score from the network representation. Indeed, it suffices to add a list of sequence numbers to nodes (or equivalently, links), representing the occurrences of the notes (or transitions from a note to the next one).

\subsection{Melodies and harmonic structures}

In the previous example, we focused on a melodic line, without considering the underlying harmonic structure of the musical piece. 
As a matter of fact, the melodic line is guided by the chord progression, since the harmonic structure implicitly influences the created melody.

It is possible to utilize the defined approach to model the harmonic structure as well. In this regard, we should note that the typical structure of a modern musical composition is based on repetitive chord progressions (with variations and modulations). Hence, the network associated with the harmonic progression would result in a simple network. 

In this work, we will focus on melodies. However, the association of melodic motifs and the corresponding chords of the harmony is another interesting aspect, to be considered in further works.

\subsection{Framework and software}

\begin{figure}
\centering
  \includegraphics[width=.7\linewidth]{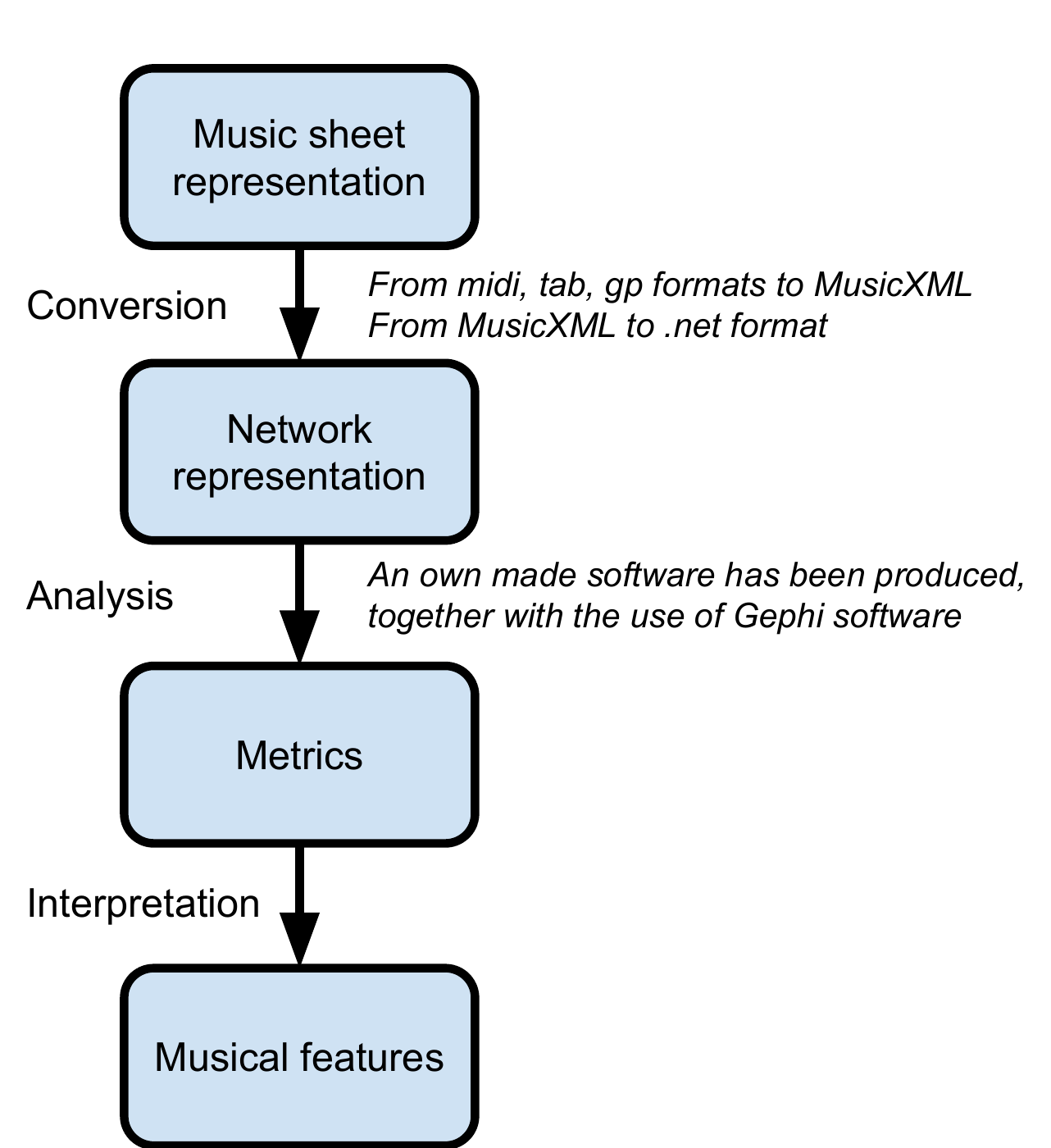}
\caption{Data workflow}
\label{fig:workflow}
\end{figure}

Figure \ref{fig:workflow} shows how the data is manipulated by the workflow process. We start from a digital representation of a music data sheet. An in-house Java and Python based conversion software has been produced that takes as input MIDI or guitar tablature file formats (e.g.,~.mid,~.tab,~.gp3,~.gp4 files). These files are converted into MusicXML documents \cite{musicXML}. Such a representation of the musical score is employed as a further input for another own made manipulation software that creates a network representation of the considered melody. This is a Java software, that in turn uses the JUNG (Java Universal Network/Graph) Framework to analyze networks and extract the metrics of interest \cite{jung}.
The Apache Commons Mathematics Library was exploited to perform the mathematics and statistics analysis \cite{apache}.
Finally, Gephi was used for the graphical representations and to measure the modularity of networks \cite{gephi}.

\begin{algorithm}[t]
\caption{Network creation process}\label{alg:creazione}
\begin{small}
\begin{algorithmic}[1]
\State $notes \leftarrow$ parse music score and extract sequence of notes
\State $prev \leftarrow$ null
\State $nodeSet \leftarrow []$, $edgeSet \leftarrow []$
\For{each $note$ in $notes$}
 \If {$note$ not in $nodeSet$}\hfill\Comment{node associated to the note not yet in the net}
  \State $nodeSet$.addNode($note$)\label{lst:line:add}
 \EndIf
 \If {$prev \neq$ null}  
      \If {$(prev, note)$ in $edgeSet$} \hfill\Comment{edge already in the net}
        \State increase weight of $(prev, note)$\label{lst:line:weight}
      \Else
        \State $edgeSet$.add($(prev, note)$) \hfill\Comment{add edge to the net}\label{lst:line:addedge}
      \EndIf
 \EndIf
 \State $prev \leftarrow note$
\EndFor
\end{algorithmic}
\end{small}
\end{algorithm}

Algorithm \ref{alg:creazione} sketches the network creation scheme utilized in the conversion process depicted in Figure \ref{fig:workflow}. The music score is parsed to extract the related set of notes, as already described.
For each note, a novel node is added if not already done (line \ref{lst:line:add}).
Then, we consider the node pair formed by the current note and the previous one; if a related edge between the two nodes already exists, its weight is increased (line \ref{lst:line:weight}); otherwise, a novel edge is added to the net (line \ref{lst:line:addedge}).

It is worth mentioning that in this study, the step from the obtained metrics to a musical interpretation, depicted in Figure \ref{fig:workflow}, is the result of a musical analysis provided by the author of this paper. The rationale was to demonstrate that certain well known musical aspects can be extrapolated from the network structure.
Nevertheless, the approach can be combined with some artificial intelligence and data mining schemes in order to obtain some more valuable and general outcomes.

\subsection{Music pieces considered in this work}

In this work, we will focus on melodies coming from different musical tracks. 
We will consider these tracks separately, showing their network representation, degree distribution and main metrics of interest. This enables a focused analysis, that allows showing several characteristics of these tracks. 

The choice was to avoid repetitive melodies, that would generate simpler networks. Hence, we selected solos of important musicians in jazz/blues/rock songs, or main melodies of classical pieces. 
In fact, in jazz and rock music, a solo allows identifying the technical artistic skills, and the ``style'' of a performer. 
More specifically, we considered the following tracks:
\begin{itemize}
 \item The solo played by Jimi Hendrix in his song titled ``Red House''. The employed instrument to play the melody was the guitar.
 \item The solo played by Miles Davis in the famous jazz piece ``So What''. The employed instrument to play the melody was a trumpet.
 \item The solo by John Coltrane in ``Giant Steps''; in this case, the instrument was a saxophone.
 \item The melodic line of the piece ``Caprice no.~24'' composed by Niccolò Paganini; the reference instrument is the violin.
 \item The melodic line of the ``Flight of the Bumblebee''. This track has a well-known melody, famous for its difficulty in terms of speed of execution. There are several versions of performers playing this tune using different instruments, e.g., piano, flute, guitar, violin, etc.
\end{itemize}
Then, when looking at the small world property, we also take into account
 \begin{itemize}
 \item The second solo played by Eric Clapton (Cream) in the song ``Crossroads'' (guitar).
 \item The solo played by  B.B.~King in ``Worried Life Blues'' (guitar).
 \item The first solo played by David Gilmour (Pink Floyd) in ``Comfortably numb'' (guitar).
\end{itemize}

\section{Metrics of Interest}\label{sec:metrics}

Complex network theory is a mathematical tool that connects the real world with theoretical research. We already mentioned that this theory is employed across a multitude of disciplines ranging from natural and physical sciences to social sciences and humanities 
\cite{Boccaletti2006175,Newman:2010}.
Thus, technological, biological, economic systems, disease pathologies, protein-protein interactions, 
can be modeled in the same way.
Focusing on multi\-me\-dia contents, it has been proved that different media, such as language and music, can be seen as a system that can be represented as a complex network 
\cite{BiemannRW12,Cancho2261,Cong2014598,icme16,Ferretti2017271,Grabska,Liu2010126,pardo}.

In this section, we introduce main metrics of interest that describe a musical track. For the interested reader, a main reference with further details on this topic and these metrics is the book by Newman on networks theory \cite{Newman:2010}.

\subsection{Length of the track}
This is the amount of notes, chords or rests composing the track. It is thus the amount of data that is used to generate the net.
This measure describes how much a performer is inclined to elaborate the melodic line he is creating.

\subsection{Number of nodes}
This measure is the total amount of nodes in a given network, that corresponds to the number of different notes that have been played during the solo, following the approach previously outlined \cite{Ferretti2017271}. This metrics gives an idea of the diversity of notes exploited to create the considered melody.

\subsection{Number of edges}
The total amount of edges in a network is the amount of connections among nodes in the network. In this specific domain, it measures how many notes have been played before/after other ones. In this case, the metrics counts the amount of existing links only, without taking into account weights.
The higher this value the higher the amount of different connections that are associated to nodes. Thus, this measure also depends on the number of nodes. 

From the number of nodes and edges, measures such as the nodes' degree and the network density can be obtained.

\subsection{Node degree and degree distribution}
The degree of a node $x$ is the amount of links that connect $x$ with other nodes in the network (included $x$ itself, if a self-loop is performed). The degree counts how many times the performer decides playing a note, after (and before) playing another one.
Being the network directed, this measure counts both links entering a node, as well as links departing from a node.
Nodes with higher degrees are notes that the artist prefers passing through in his melodies, meaning that the considered note is played before (after) a high number of other ones.

Given the whole nodes' degree, it is possible to build a degree distribution, stating how much notes are connected in general.
By counting how many nodes have each degree, we form the degree distribution $P(k)$, defined by
$$P(k)=\text{fraction of nodes in the network with degree }k.$$

In most cases, insights can be obtained by plotting the degree distribution, both in linear and log-log scales. The linear scale allows understanding if there is a wide variability on the nodes' degrees, if some nodes are \emph{hubs}, meaning that these nodes have a large amount of links with other nodes, or rather if nodes have similar degrees.

Under the circumstance that most nodes have a relatively small degree, but a few nodes have very large degree (hubs), being connected to many other nodes, it is interesting to plot the degree distribution in a log-log scale. This allows assessing if the degree distribution follows a power law distribution, i.e., the probability that a node has a degree $k$ is $P(k) \sim k^{-\lambda}$, for some positive value $\lambda$.
In fact, it suffices to assess if, in a scatter plot on log-log scale of the degree distribution, points lie approximately along a line. 
Networks with power-law distributions are called \emph{scale-free}, since power law distributions have the same functional form at all scales \cite{Newman:2010}. The probability $P(k)$ remains unchanged (other than a multiplicative factor) when rescaling the independent variable $k$, as it satisfies $P(ak)=a^{-\lambda} P(k)$. In simpler words, scale invariance means that the overall features of our network look the same (at least statistically) under dilatations; if we take a degree distribution and zoom in a given portion of it, we would notice that the distribution in that portion has the same trend of the original one, no matter how much we zoom in (similarly to what happens in fractals \cite{ref_nonsense5,ref_nonsense4}).

\subsection{Density}
Network density measures how close the network is to be a completely connected net. In other words, a complete graph has all possible edges among nodes and its density would be equal to 1.
The network density is thus measured as 
$$\text{density} = \frac{\#\text{edges}}{\text{potential connections}}$$
where the number of potential connections, in directed nets with self-loops, is $\#\text{nodes}^2$.

\subsection{Average distance}
Average distance is the average path length needed to go from one node to another one in the network. This metrics gives an idea of how complex the solo is. In fact, larger networks might have higher distances. However, higher average distances mean that the player is used to move ``locally'' among notes he usually plays together and that, going from one note to another one, a high amount of notes should be traversed, on average. Thus, in the case of musical tracks, this suggests that the player is used to play ``near'' notes, i.e.,~there is a preference to combine certain groups of notes to create the melody.
Indeed, the presence of short paths is a revealing factor to assess if a network is a small world (as discussed in the remainder of the paper), and this measure should be considered together with the clustering coefficient and the network density.

Distances can be measured using the directed graph, as well as the undirected version of the graph, obtained by removing the direction information and considering links as bidirectional ones. Clearly enough, distances obtained for the undirected networks are usually lower than those of the corresponding directed ones.
Implementations of classic algorithms to compute the distance measures in networks are available in many software tools for network analysis, such as those employed in this work, i.e., Gephi \cite{gephi}, Jung \cite{jung}.

\subsection{Diameter}
The diameter is the maximum average distance in a network. Also this measure can provide some insights on the complexity of the melody, similarly to the average distance.

\subsection{Clustering coefficient} 
The clustering coefficient is a measure assessing how much nodes in a graph tend to cluster together.
Let consider a node $x$ and the neighborhood of $x$, i.e.,~the set of nodes that are connected to $x$. In a clustered network, there is a high probability that a node in the neighborhood of $x$ is connected to other nodes in the neighborhood of $x$ \cite{Newman:2010}.
Thus, for instance, in social networks the clustering coefficient measures to what extent friends of a node are friends of one another too.

A common way to assess if a network has a high clustering coefficient is to check for the presence of triangles in the network, i.e., given two links $(x,y)$, $(x,z)$, sharing the node $x$, then it is likely that a third link $(y,z)$ exists such that the three links form a triangle.
Indeed, the clustering coefficient of a net is defined as
$$C = \frac{3\times \#\text{triangles}}{\#\text{connected triplets}} = \frac{\#\text{closed triplets}}{\#\text{connected triplets}}$$
where we are considering triplets of connected nodes and triangles (or closed triplets) formed by nodes.

In this domain, the clustering coefficient states how much notes are clustered, i.e., how much the performer plays notes in an interchangeable order, since there are triplets of notes that are grouped as triangles in the network.

\subsection{Betweenness} 
Betweenness is a centrality measure that indicates if a node has a large influence in the network \cite{Ferretti2017271}. It basically measures how often one must pass through a given node going from an origin to a destination. 
Thus, betweenness identifies those notes the player/composer prefers passing through in his solo/melody.

Betweenness of a node $x$ is defined as 
$$\text{bet}(x) = \sum_{y \neq x \neq z} \frac{\sigma_{yz}(x)}{\sigma_{yz}},$$ 
where $\sigma_{yz}$ is the total amount of shortest paths in the network going from $y$ to $z$, and $\sigma_{yz}(x)$ is the number of those paths passing through $x$ \cite{Ferretti2017271,Newman:2010}.

\subsection{Modularity and community detection} 

Modularity measures how well a network decomposes into communities. In other words, it determines if the network can be grouped into sets of nodes, which are densely connected internally.
The modularity of a partition is a scalar value that measures the density of links inside communities as compared to links between communities.
A high modularity score indicates a sophisticated internal structure, i.e., how the network is compartmentalized into sub-networks.
In the case of weighted networks, it is defined as \cite{1742-5468-2008-10-P10008}
$$Q = \frac{1}{2m}\sum_{ij}\left( w_{ij}-\frac{k_i k_j}{2m} \right) \delta(c_i,c_j)$$
where 
$w_{ij}$ is the weight of the edge $(i,j)$;
$k_i$ is the sum of all the weights of edges attached to node $i$;
$c_i$ is the community to which node $i$ belongs;
$\delta()$ is the delta function, i.e.,~$\delta(u,v)$ is $1$ if $u=v$, $0$ otherwise;
and $m=\frac{1}{2}\sum_{ij}w_{ij}$.
From this definition, different implementation algorithms can be devised. In this work we employed the implementation provided in Gephi \cite{gephi}, that follows the algorithm presented in \cite{1742-5468-2008-10-P10008}.

In musicological terms, the presence of communities would indicate that the musician is inclined to work with specific groups of notes at a time, in a given melody.

\section{Practical Examples}\label{sec:ex}

\subsection{Analysis of some tracks}

\begin{figure}
\centering
\begin{minipage}{0.6\textwidth}
   \centering
  \includegraphics[width=\linewidth]{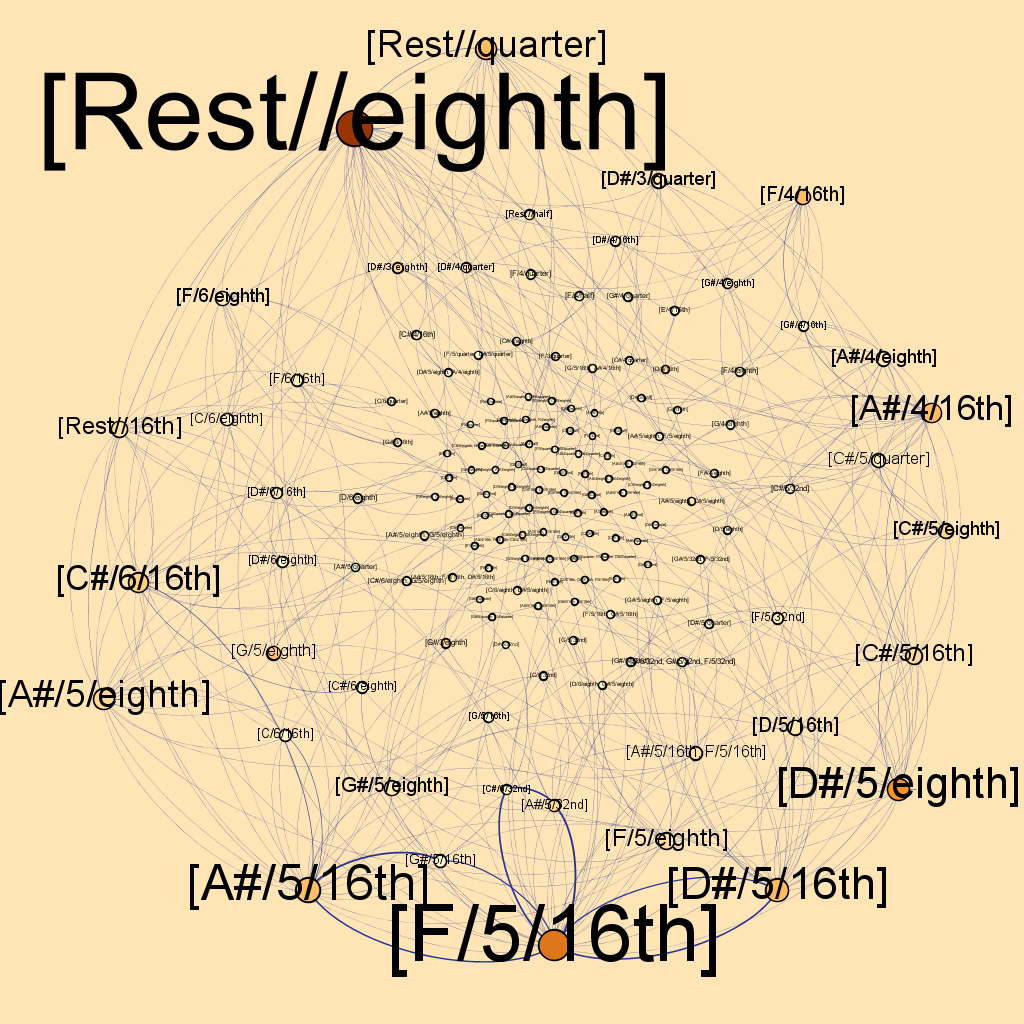}
\end{minipage}%
\begin{minipage}{0.4\textwidth}
   \centering
  \includegraphics[width=\linewidth]{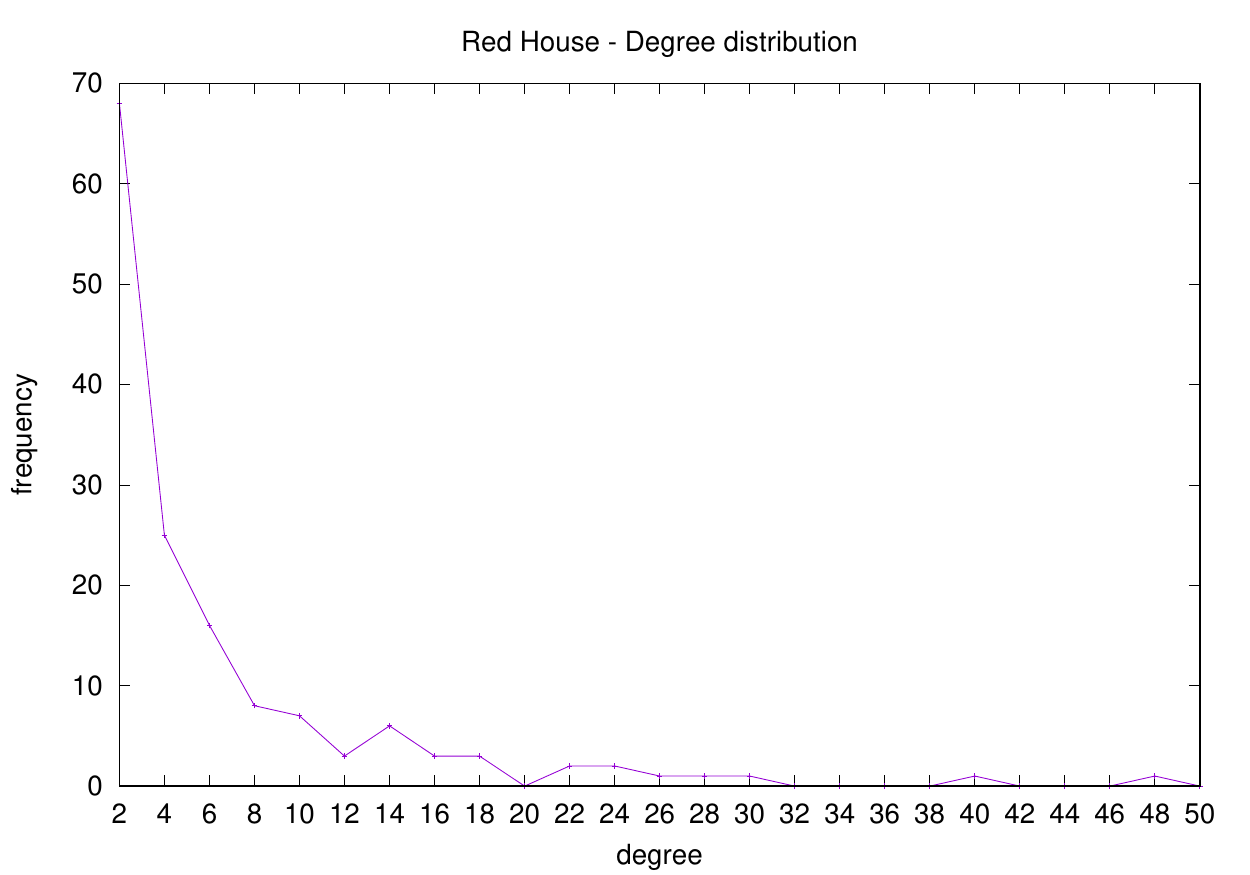}
  \includegraphics[width=\linewidth]{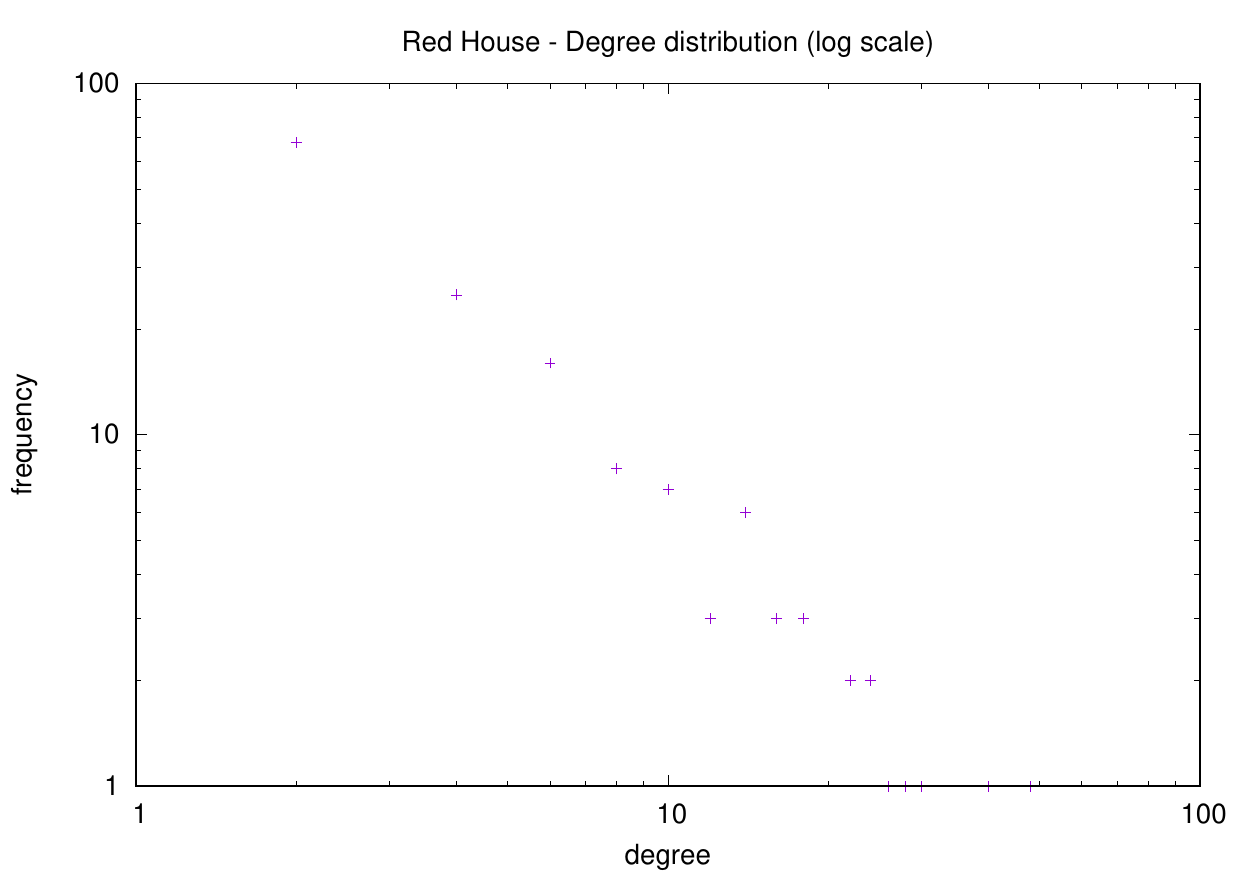}
\end{minipage}%
\caption{Jimi Hendrix -- Red House}
\label{fig:hendrix}
\end{figure}

\begin{figure}
\centering
\begin{minipage}{0.6\textwidth}
   \centering
  \includegraphics[width=\linewidth]{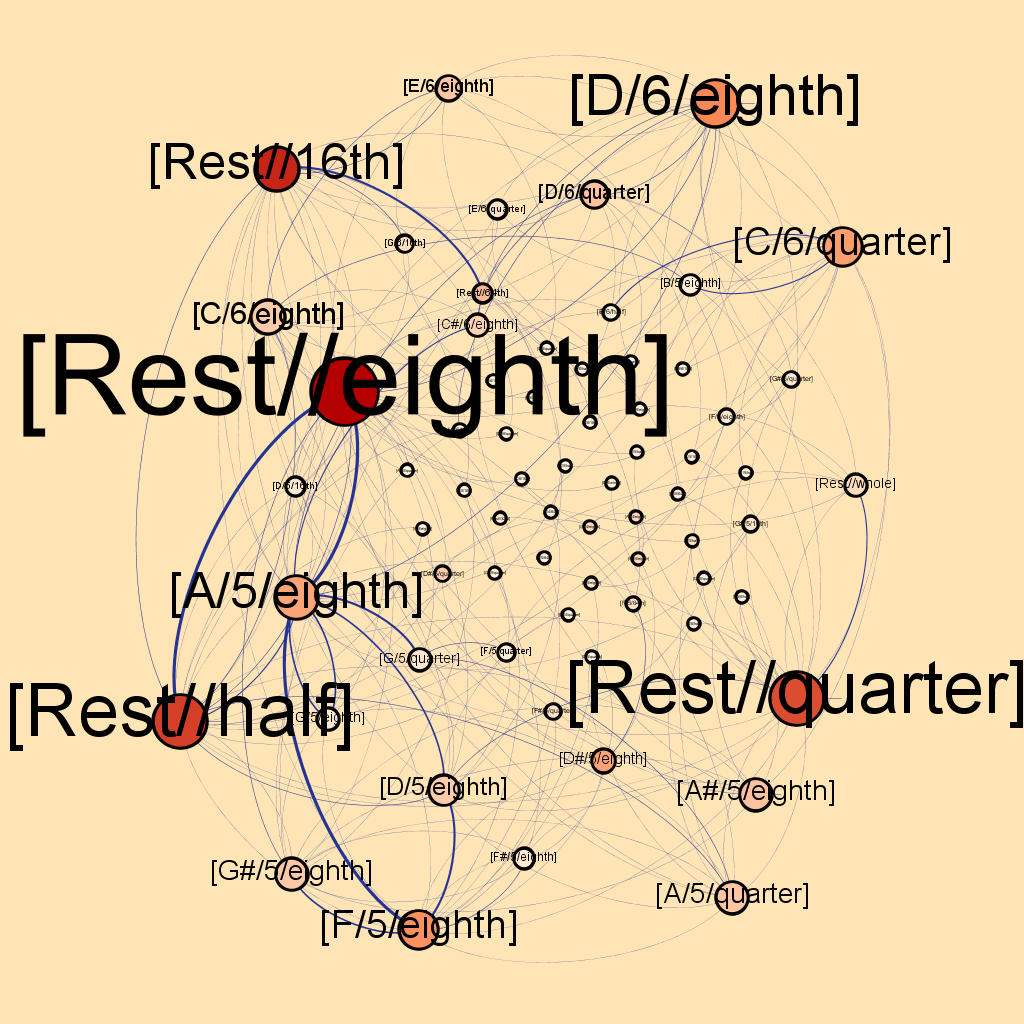}
\end{minipage}%
\begin{minipage}{0.4\textwidth}
   \centering
  \includegraphics[width=\linewidth]{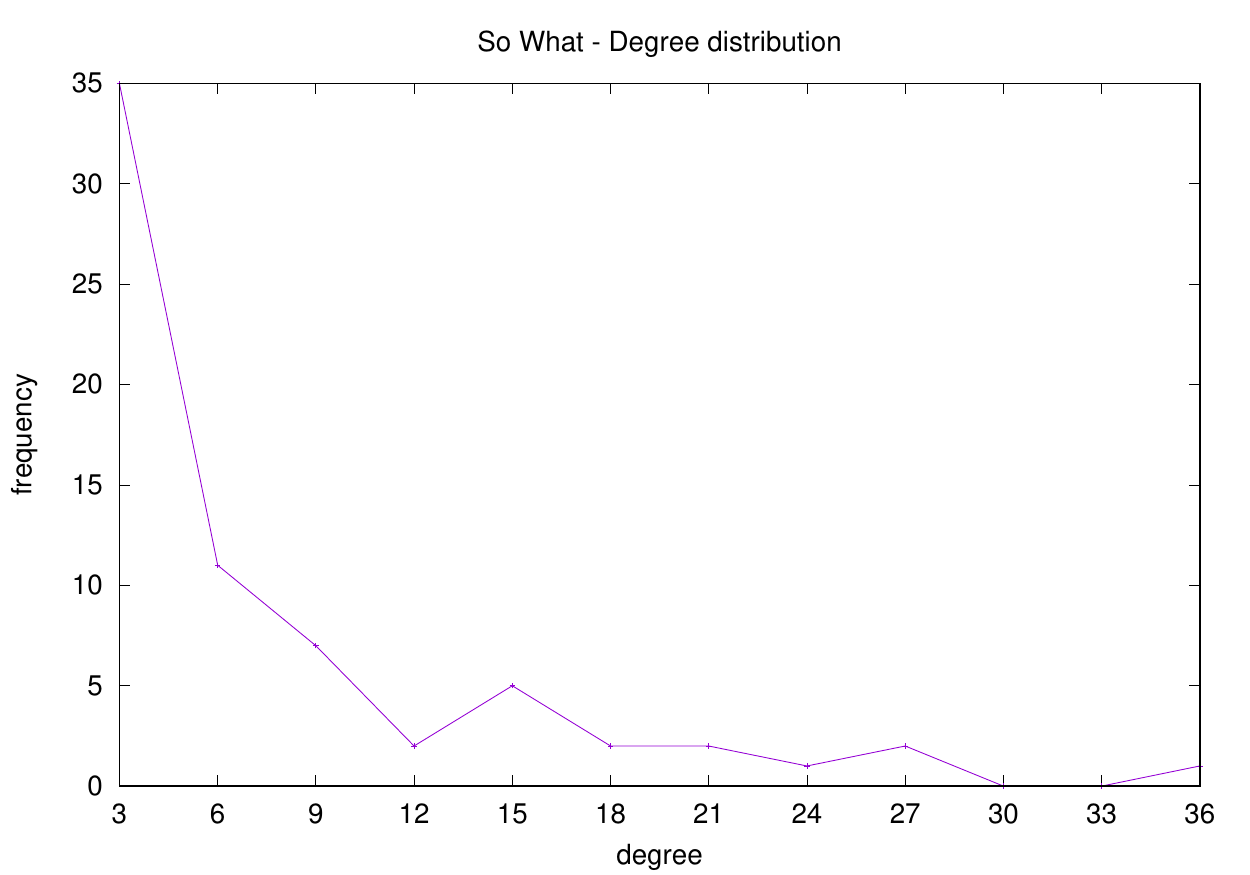}
  \includegraphics[width=\linewidth]{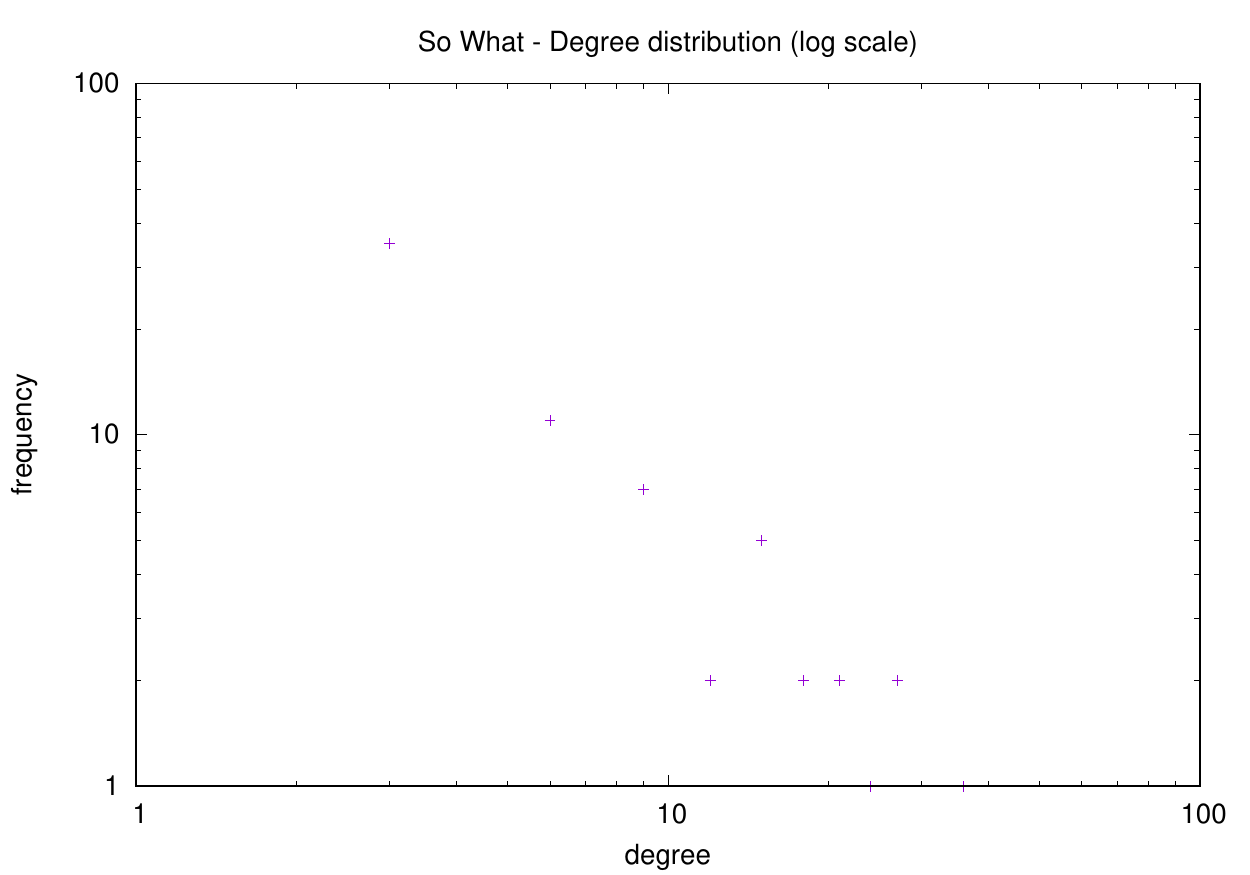}
\end{minipage}%
  \caption{Miles Davis -- So What}
  \label{fig:davis}
\end{figure}

\begin{figure}
\centering
\begin{minipage}{0.6\textwidth}
   \centering
  \includegraphics[width=\linewidth]{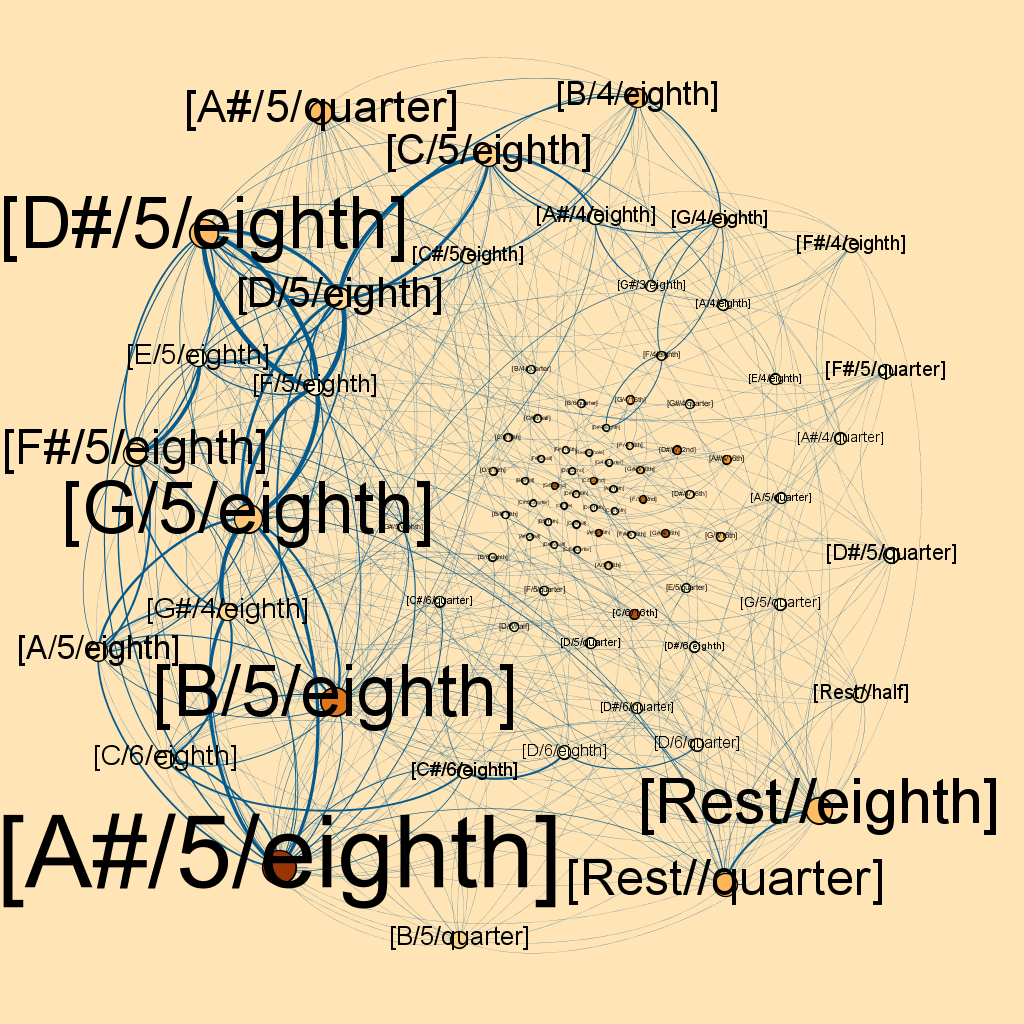}
\end{minipage}%
\begin{minipage}{0.4\textwidth}
   \centering
  \includegraphics[width=\linewidth]{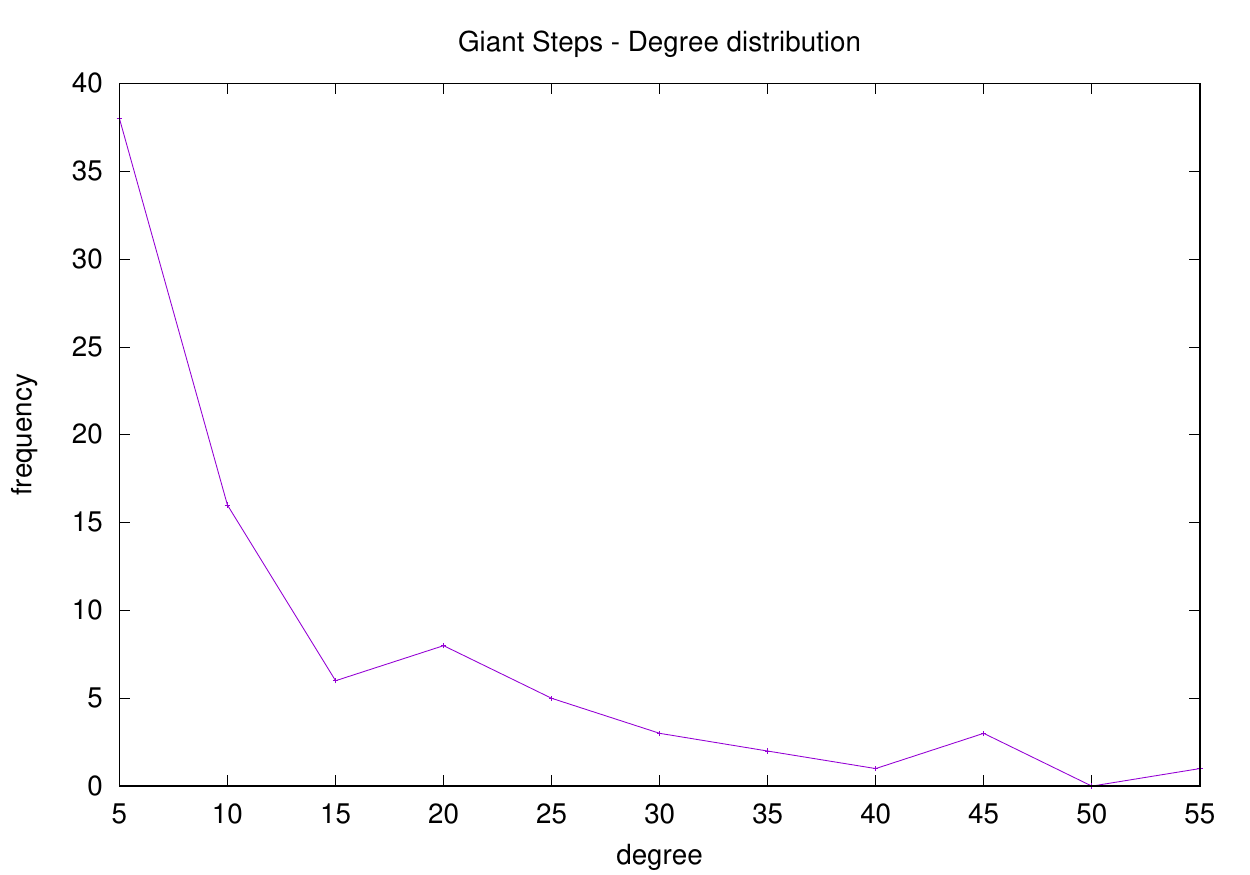}
  \includegraphics[width=\linewidth]{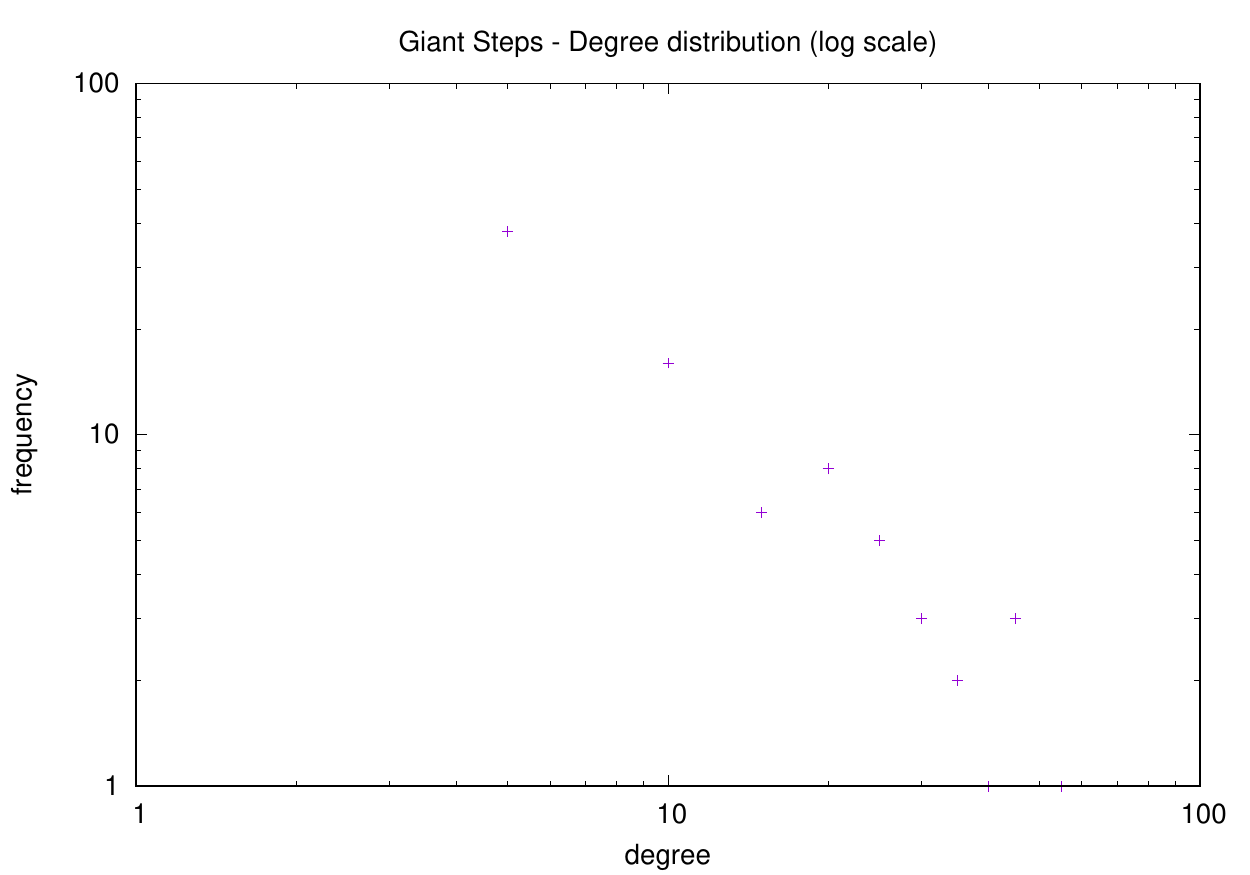}
\end{minipage}%
  \caption{John Coltrane -- Giant Steps}
  \label{fig:coltrane}
\end{figure}

\begin{figure}
\centering
\begin{minipage}{0.6\textwidth}
   \centering
   \includegraphics[width=\linewidth]{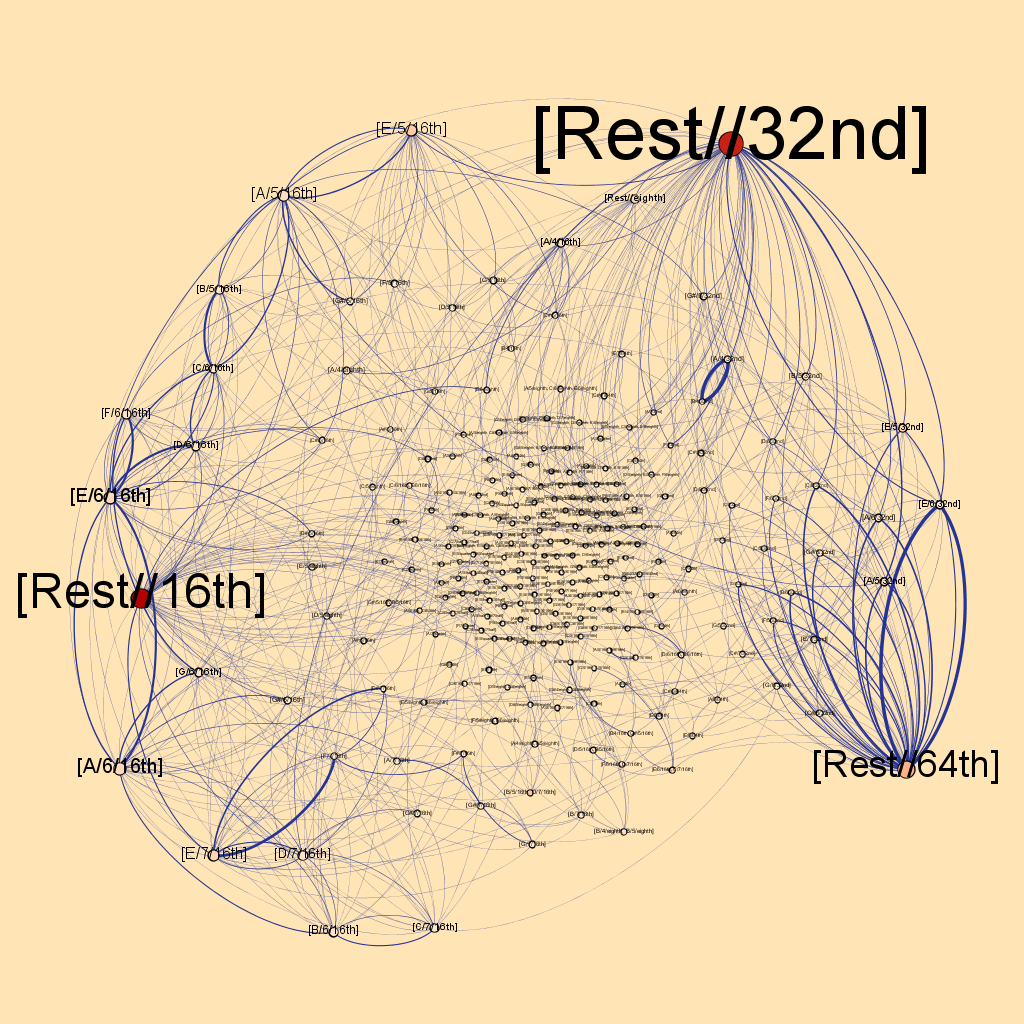}
\end{minipage}%
\begin{minipage}{0.4\textwidth}
   \centering
  \includegraphics[width=\linewidth]{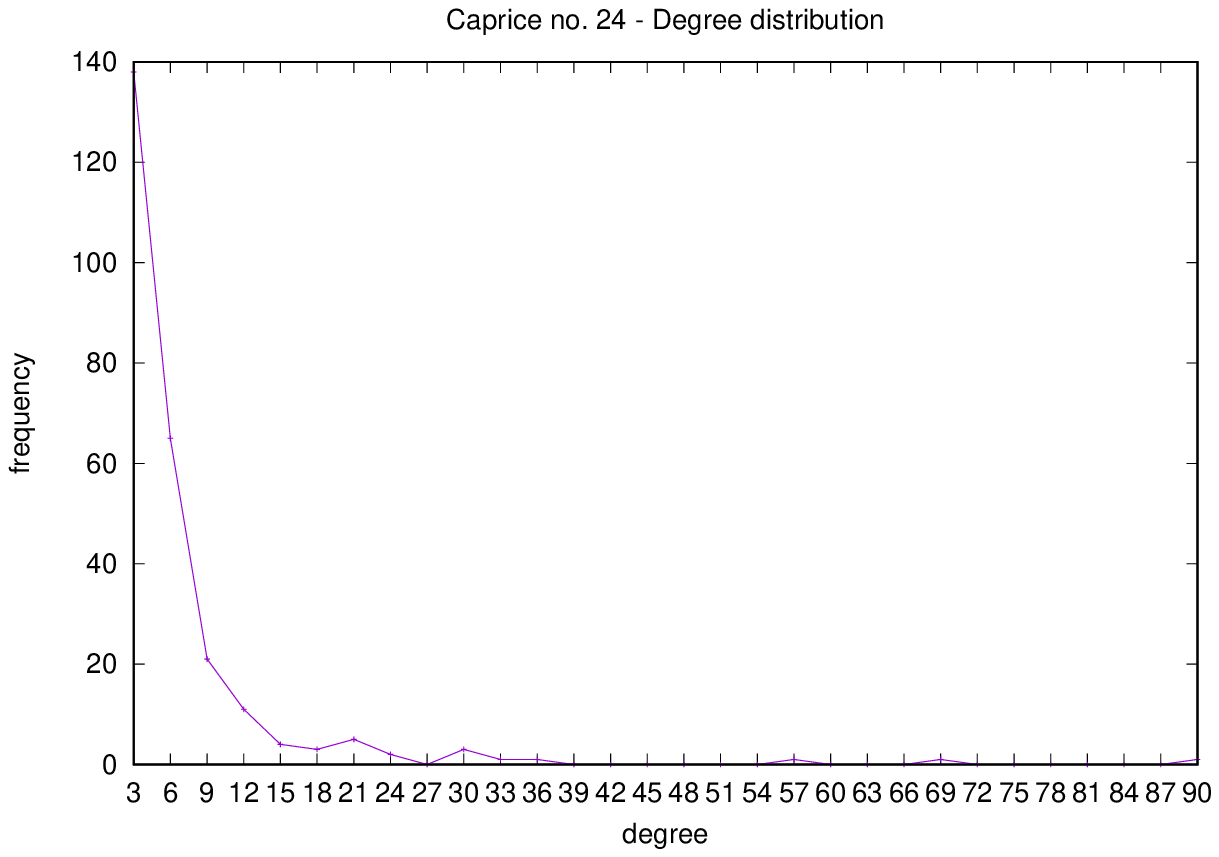}
  \includegraphics[width=\linewidth]{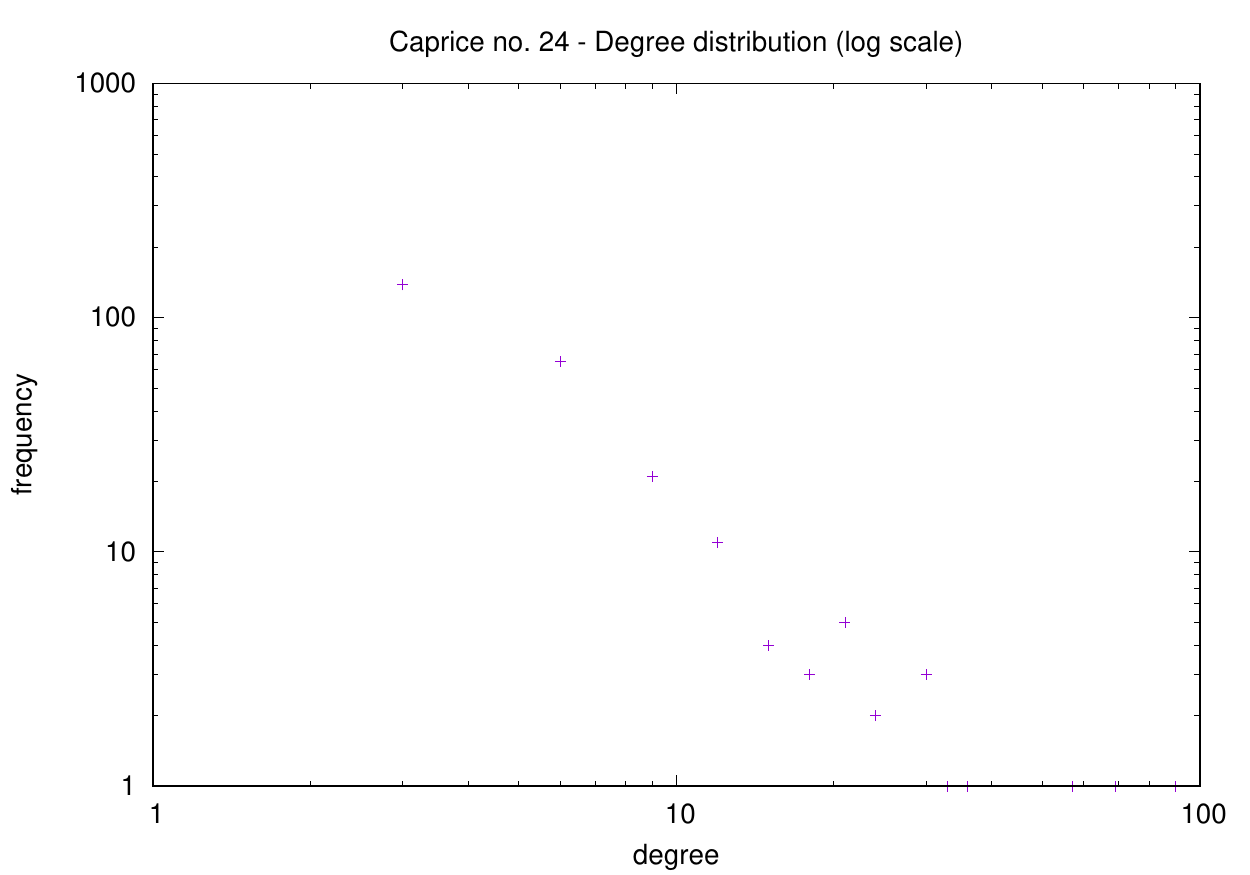}
\end{minipage}%
  \caption{Niccolò Paganini -- Caprice no.~24}
  \label{fig:paganini}
\end{figure}

\begin{figure}
\centering
\begin{minipage}{0.6\textwidth}
   \centering
   \includegraphics[width=\linewidth]{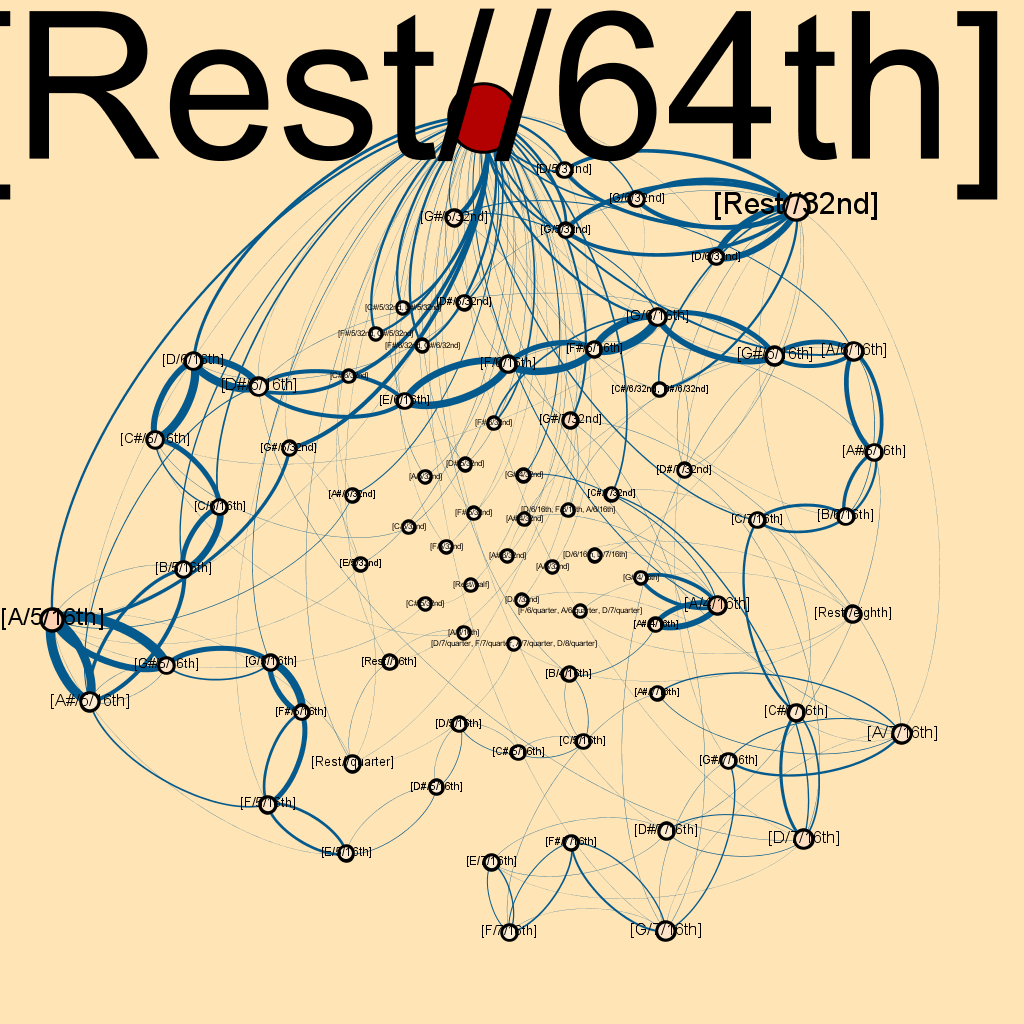}
\end{minipage}%
\begin{minipage}{0.4\textwidth}
   \centering
  \includegraphics[width=\linewidth]{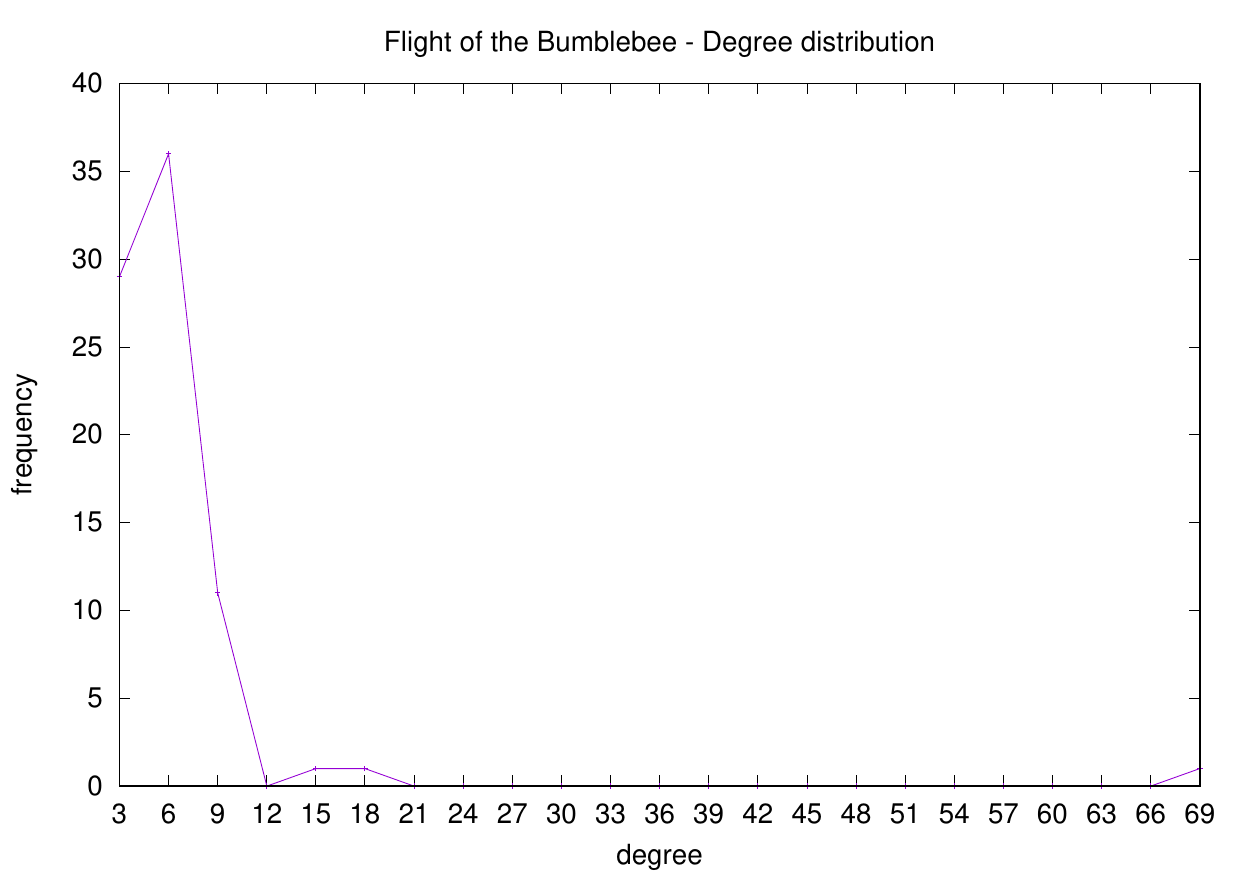}
  \includegraphics[width=\linewidth]{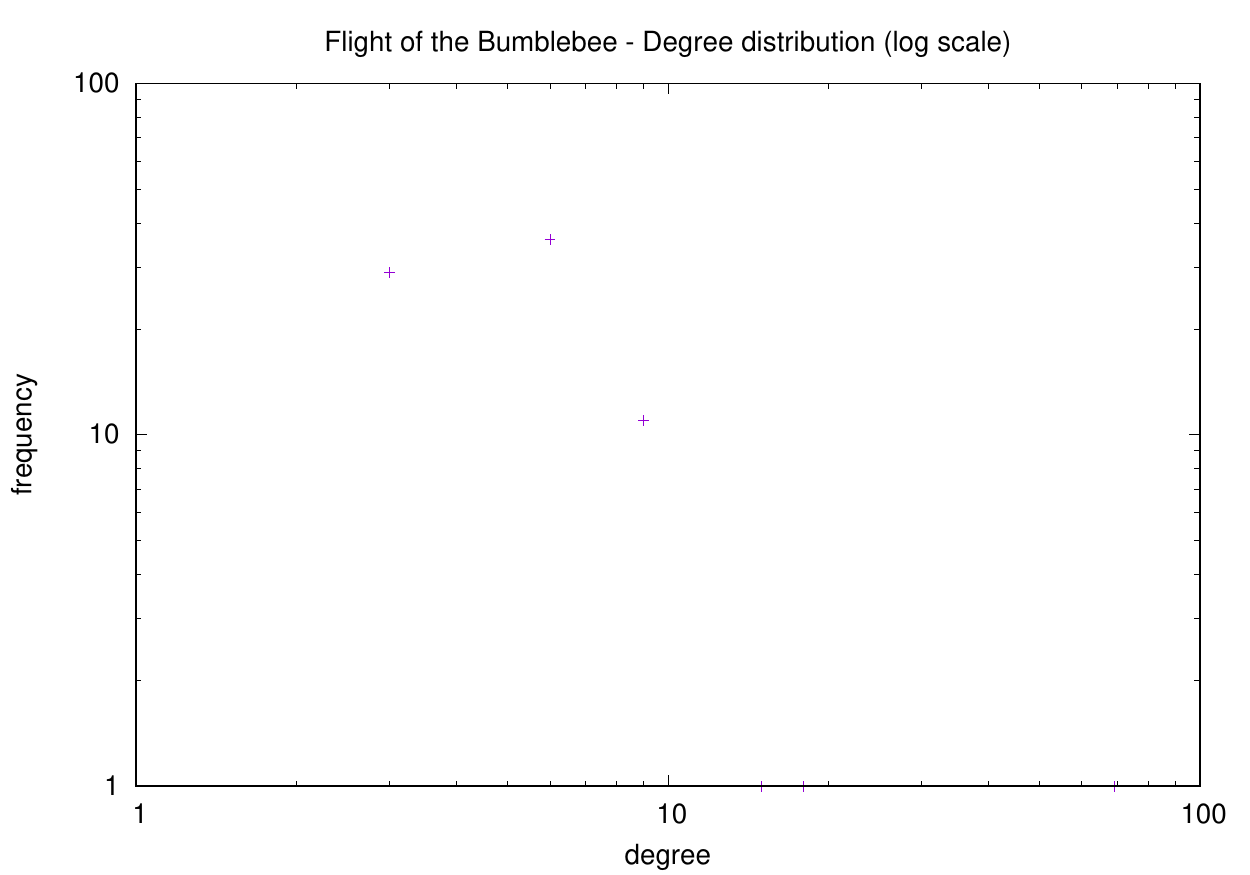}
\end{minipage}%
  \caption{Nikolai Rimsky-Korsakov -- Flight of the Bumblebee}
  \label{fig:flight}
\end{figure}

Figures \ref{fig:hendrix}--\ref{fig:flight} show examples of the network representations of different melody lines, together with their associated degree distribution, plotted in linear and log-log scales (charts on the right side). 
In particular, Figure \ref{fig:hendrix} shows the network of the solo played by Jimi Hendrix in ``Red House''.
Figure \ref{fig:davis} shows the solo played by Miles Davis in ``So What''.
Figure \ref{fig:coltrane} reports the network of the solo by John Coltrane in ``Giant Steps''.
Figure \ref{fig:paganini} is the network of the piece ``Caprice no.~24'' composed by Niccolò Paganini.
Figure \ref{fig:flight} is the network associated to the melodic line of the ``Flight of the Bumblebee''. 

In each network, nodes have labels associated of the form ``[pitch/octave/du\-ra\-tion]'' where, as the names say, ``pitch'' is the pitch of the sound, ``octave'' is a number counting the octave of the note, and ``duration'' specifies the relative duration of the note, with respect to a given bar\footnote{A bar (often referred as measure) is a segment of time corresponding to a specific number of beats in which notes are played. Dividing music into bars provides regular reference points to pinpoint locations within a piece of music.}.

In the figures, nodes have a size (and label font) that is proportional to the degree of the node, i.e., the higher the amount of connections the larger the node (and label) size. The color of nodes, instead, is proportional to the measure of the betweenness centrality measure; the more the color goes to red the higher the betweenness of that node.
Thus, nodes with a red color and a larger size reflect important, common notes in the melodic line.

Links are depicted based on their weight, i.e., the larger the line the higher the weight of that link in the network, meaning that the notes pair connected through that link has been played multiple times. 

  
By looking at the figures, it is clear that rests play an important role in every song (or solo). If we look at nodes with highest degrees in every depicted network, we will find some rest nodes in there.

Some preliminary insights on musical features can be captured by a first look of the network. However, a more detailed and deeper analysis would allow to extrapolate more articulated characteristics.
We can find some interesting elements by looking at
Table \ref{tab:analisi}, which reports measurements of some main metrics related to networks associated to the tracks considered in Figures \ref{fig:hendrix}--\ref{fig:flight}. 
For each network, the table shows
the number of network nodes, number of edges, length of the considered track, average degree of the net (``avg deg''), maximum degree (``max deg''), median degree (``median deg''), network diameter (``diam''), average clustering coefficient (``cc''), average path distance ($L$) and network density.

As already mentioned, the number of network nodes corresponds to the amount of different notes played by a performer in each song melody/solo. Notes of different durations, played at different octaves account for different nodes. Similarly, chords account for single nodes, different to those associated the the single notes composing each chord. For this reason, the amount of nodes results quite larger than the twelve notes, labeled in the Western tonal music.

The number of edges, the measures concerned with the degrees and the network density assess how much the considered melody/solo has interconnections among notes (the larger values the more varied the use of notes in the melodies). 
The diameter and the average path distance $L$ are measures concerned with distances among nodes. Hence, they give an idea of how many notes are to be played going from a given note to another, also with respect to the length of the melody. Finally, as mentioned the clustering coefficient states how much notes are played in an interchangeable order and how much they are grouped in musical phrases.

\begin{landscape}
    \topskip0pt
    \vspace*{\fill}
\begin{table*}[th]
\caption{Some metrics from the network analysis: for each song the reported metrics are the number of network nodes, number of edges, length of the considered track, average degree of the net (``avg deg''), maximum degree (``max deg''), median degree (``median deg''), network diameter (``diam''), average clustering coefficient (``cc''), average path distance ($L$), network density.}
\label{tab:analisi}
\scriptsize
\begin{tabular}{|| l || c | c | c | c | c | c | c | c | c | c ||}
  \hline			
  \hline			
  track &                                       \#nodes  & \#edges & length   & avg deg  & max deg & median deg   & diam     & cc         & $L$ & density\\
  \hline  
  \hline			
   J.~Hendrix -- Red House                       & 148   & 458     & 802      & 6.19     & 47      & 4            & 15       & 0.16       & 4.5    & 0.02 \\
   \hline			
   M.~Davis -- So What                           & 68    & 230     & 305      & 6.77     & 35      & 3            & 7        & 0.21       & 3.18   & 0.05 \\
   \hline		
   J.~Coltrane -- Giant Steps                    & 83    & 459     & 1507     & 11       & 51      & 6            & 16       & 0.3        & 4.1    & 0.07 \\
   \hline			
   N.~Paganini -- Caprice no.~24                 & 257   & 754     & 1331     & 5.87     & 89      & 3            & 22       & 0.14       & 5.24   & 0.01 \\
   \hline			
   N.~Rimsky-Korsakov -- Flight of the Bumblebee & 79    & 215     & 1393     & 5.44     & 68      & 4            & 11       & 0.14       & 3.65   & 0.04 \\ 
  \hline  
  \hline			
\end{tabular}
 \end{table*}
    \vspace*{\fill}
\end{landscape}

The solo by Jimi Hendrix in ``Red House'' (Figure \ref{fig:hendrix}) denotes a certain complexity of the network, with prominent notes having a pitch among \emph{A\#, D\#, F}. Indeed, the song is a blues played in the \emph{A\#} key and these three notes are the tonic\footnote{In music, the ``tonic'' is the first scale degree of a diatonic scale. It is thus the tonal center of a given key; in other words, it is the main note of that key.} of the main song chords.
The charts related to the degree distribution witness a wide variability on the nodes' degrees. In particular, the median degree is 4, the average degree is around 6, while the highest degree is 47 (as reported in Table \ref{tab:analisi}). The log-log scale chart in Figure \ref{fig:hendrix} shows that degrees lie approximately along a line; thus, the network is a scale-free.
The network is composed of a relatively high number of nodes (with respect to other considered network exemplars); this means that there might be several notes with the same pitch, but different durations. The higher number of nodes corresponds to a lower density than other nets (apart from ``Caprice no.~24'', which is the largest considered net) and a high diameter (Table \ref{tab:analisi}). However, the average degree and average path distance are low, and the clustering coefficient has a significant (but not excessive) value. 
These values might be explained if we consider that we are dealing with a classic electric blues guitar player, playing over a classic blues. It is well known in musicology that (non-jazzy) blues players employ a simple underlying chord structure, and the associated reference scales are limited in number. Nevertheless, the melody is quite intricate and syncopated. Moreover, the player here employs a high number of bi-chords (i.e., two notes played simultaneously), that increase the final number of nodes in the network.

The solo by Miles Davis in ``So What'' (Figure \ref{fig:davis}) has a simpler structure with respect to other networks. Rests are largely employed and are central in the network (high betweenness). Indeed, music experts quite often discuss on the ability of Miles Davis to use silence periods in his solos. His famous quotation ``\emph{It's not the notes you play, it's the notes you don't play}'' surely confirms this claim.
According to the degree distributions, despite the lower amount of nodes, the degree distribution seems to follow a power law, making the net a scale-free.
In Table \ref{tab:analisi} it is possible to appreciate that this simple structure is reflected on the reported metrics, i.e.,~lower number of nodes, edges, average distance, maximum and median degrees. These values are mainly due by the shorter length of this solo. The low amount of nodes corresponds to a higher clustering coefficient and network density, meaning that the player combines these nodes in different orders during his solo.

``Giant Steps'' solo is quite complex (Figure \ref{fig:coltrane}); many music excerpts have been written to analyze this solo. The network reveals this complexity. The harmonic structure of the track is based on three main keys, i.e.,~$D\#, G, B$.
Important nodes in the network are related to the tonic notes 
of these keys, or the dominant notes\footnote{In music, the ``dominant'' note in a given key is the fifth scale degree of the diatonic scale: It is called dominant because it is next in importance to the tonic.}, i.e., $A\#$ for $D\#$, $F\#$ for $B$.
Also this network is a scale-free one, as confirmed by the shape of the degree distribution in Figure \ref{fig:coltrane} (left charts).
Metrics reported in Table \ref{tab:analisi} provide some important insights. In fact, while the solo of ``Giant Steps'' has a high length w.r.t.~other considered exemplars, its number of nodes is relatively lower. This means that, for instance, in ``Giant Steps'' the amount of nodes (i.e., notes with a given pitch and duration) played by Coltrane is lower than the number of (nodes, and thus) notes played by Hendrix in ``Red House'' (which has a lower length also). All this, despite the fact that the solo in ``Giant Steps'' is considered as a particularly complicated one, due to the complex underlying harmony and the complex melody that Coltrane builds on top of it. Anyway, the amount of edges is the same (one unit higher, actually) of ``Red House''. This means that these notes are widely interconnected, and this results in an intricate and complex melody. 
Indeed, the complexity of this track is confirmed by the highest average degree, high median degree, the high diameter, average path length and network density.

``Caprice no.~24'' has the highest amount of nodes and number of edges, with a limited average degree. In fact, apart from some rests, no specific notes appear to have a main role, with respect to others (Figure \ref{fig:paganini}). In this case, rests have the highest degrees and betweenness values. The net has a scale-free structure. As shown in Table \ref{tab:analisi}, the average degree is lower than other considered networks, as well as the median degree and the average distance. The fact that this net shows the highest maximum degree (w.r.t.~other networks) witnesses the importance of the hubs, which are indeed rest notes.
The diameter is quite high, due to the largest network size; this is confirmed by the low network density.
In substance, the importance of rests, the low median degree, network density and clustering coefficient confirm a linear structure of this classical composition.

As concerns ``The Flight of the Bumblebee'', the main role of a very short rest (rest 64th) is evident from Figure \ref{fig:flight}. The corresponding node has the highest degree and highest betweenness value. Moreover, the network reveals a presence of recurrent sequences of note pairs. There are several links with high weights, and these note pairs appear to be organized as a chain.
Indeed, this track is characterized by repetitions of chromatic sequences of notes, and this is confirmed by the network structure.
In this case, the degree distribution suggests that this network does not follow a power law; thus, the net does not show a scale-free property.
All these considerations are confirmed by the values in Table \ref{tab:analisi}. In fact, despite the long duration of the track, the network has a low number of nodes, edges, but a not so high diameter, due to the fact that the ``rest 64th'' node plays the role of hub, connecting different portions of the network.
The long duration of the track (length) and the high weights of the edges confirm (for those that do have a musical knowledge of the track) that the track has a repetitive structure in the melody.


\begin{table*}[th]
\centering
\caption{Small world property: comparison between the clustering coefficient (column ``cc'') and the average distance (column $L$) of the considered \textbf{undirected} network, with the clustering coefficient (column ``cc (RG)'') and the average distance (column $L_{RG}$) of the corresponding random graph. The last column shows the small coefficient as defined in Equation \ref{eq:sigma}; basically, the small property exists when $\sigma\gg1$.}
\label{tab:sw}
\scriptsize
\begin{tabular}{|| l || c | c | c | c | c |c ||}
  \hline			
  \hline			
  track & cc & cc (RG) & $L$ & $L_{RG}$ & $\sigma$ \\
  \hline  
  \hline			
  J.~Hendrix -- Red House & 0.24 & 0.02 &	3.37 &	5.00 & 17.8 \\
  \hline			
  M.~Davis -- So What & 0.32 & 0.05 & 2.51 & 4.22 & 10.7\\
  \hline		
  J.~Coltrane -- Giant Steps & 0.40 & 0.07 & 4.56 & 4.42 & 5.5 \\
  \hline			
  N.~Paganini -- Caprice no.~24 & 0.20 & 0.01 & 3.71 & 5.55 & 29.9 \\
  \hline			
  N.~Rimsky-Korsakov -- Flight of the Bumblebee & 0.25 & 0.03 & 2.60 & 4.37 & 14\\
  \hline			
  E.~Clapton (Cream) -- Crossroads (2nd solo) & 0.40 &	0.04 &	3.68 &	4.29 & 11.6 \\
  \hline			
  B.B.~King -- Worried Life Blues & 0.09 & 0.05 & 3.04 & 3.58 & 2.1 \\
  \hline			
  Pink Floyd -- Comfortably numb (1st solo) & 0.06 & 0.03 & 4.30 &	4.03 & 1.9 \\
  \hline  
  \hline			
\end{tabular}
\end{table*}

\subsection{Small world property}

Table \ref{tab:sw} shows an analysis of some specific solos, aimed at assessing if the related networks exhibit a small word property. 
In simple words, when a network is identified as a small world, we can conclude that the related solo is composed of a sequence of nodes which are combined and played in various orders (i.e., we can identify subsets of nodes in the net that are highly interconnected among themselves), with a significant amount of connections between notes that are in different clusters (or, in some sense, in different ``areas'' of the network) \cite{Ferretti2017271}.
A small world is a highly clustered net with a small average path length. 

To mathematically assess these features, a method is to compare the network against a graph of the same size, where node links are randomly generated \cite{Newman:2010}.
In this case, in order to be a ``small world'' the network should have a (small) average distance comparable to that of the considered random graph, but a significantly higher clustering coefficient.
In particular, if one looks at the clustering coefficient ($cc$) together with the average distance ($L$) of the considered network (note than in this case we ignore the directed nature of the links, thus obtaining an undirected network \cite{Newman:2010}), and the clustering coefficient ($cc_{RG}$) together with the average distance ($L_{RG}$) of the corresponding random graph, we can measure the small-coefficient as
\begin{equation}\label{eq:sigma}
\sigma=\frac{cc/cc_{RG}}{L/L_{RG}},\end{equation}
concluding that the network can be classified as a small world when $\sigma$ is significantly higher than $1$ \cite{journal.pone.0002051}.

From Table \ref{tab:sw}, we can appreciate that all the considered solos have a corresponding value of the coefficient $\sigma$ higher than $1$; thus, in general they might be considered as small worlds. 
As a matter of fact, a comparison has been made on musical solos of a set of different guitar players \cite{Ferretti2017271}.
From the database considered in \cite{Ferretti2017271}, we measured the value of $\sigma$ and found that all these solos feature a value $\sigma > 1$. 
However, while the majority of solos has a $\sigma$ significantly higher than $1$, others exist having a $\sigma$ value near $1$. Two examples, reported in Table \ref{tab:sw} (two last rows), are the solo played by D.~Gilmour (Pink Floyd) in ``Comfortably numb (1st solo)'' and ``Worried Life Blues'' by B.B.~King. In fact, these two solos are poorly clustered with respect to other examples.
Indeed, the corresponding music melodies can be considered as linear, ``melodic'' ones. Thus, in these two cases a small world property is not fully evident.

\subsection{Modularity}

We already mentioned that through the measure of modularity we can understand if it is possible to decompose a network into communities.
This would indicate that the performer is inclined to work with specific groups of notes at a time, in a given melody.

To better understand if and when melodies can be partitioned into subsets of notes, we considered the networks already described, but rest notes were removed from the network. In fact, rests have the role of separating musical notes and their removal enhances the partitioning of communities. Indeed, after this removal, in certain cases some notes communities are not connected to the rest of the network, meaning that they were played within music intervals delimited by rests.
Table \ref{tab:modularity} shows the modularity value for these networks, measured using the approach described in \cite{1742-5468-2008-10-P10008} and implemented in Gephi \cite{gephi}, together with the amount of identified communities. The higher the modularity value the more evident the presence of communities in each network. In general, we can see that a community-based structure is more evident in classical tracks, i.e., Caprice no.~24 and the Flight of the Bumblebee, where modularity values are higher, as well as the amount of communities (even in these two cases some communities are related to isolated notes). Conversely, it is more difficult to divide the network into communities for the jazz solo in Giant Steps. 

\begin{table*}[th]
\centering
\caption{Modularity: for each track the modularity measure is reported (the higher the value the more evident the presence of communities), together with the amount of communities. Rests have been removed from the tracks.}
\label{tab:modularity}
\small
\begin{tabular}{|| l || c | c ||}
  \hline			
  \hline			
  track & modularity & \# communities \\
  \hline  
  \hline			
  J.~Hendrix -- Red House & 0.49 & 9 \\
  \hline			
  M.~Davis -- So What & 0.46 & 12\\
  \hline		
  J.~Coltrane -- Giant Steps & 0.35 & 5 \\
  \hline
  N.~Paganini -- Caprice no.~24 & 0.62 & 23 \\
  \hline
  N.~Rimsky-Korsakov -- Flight of the Bumblebee & 0.68 & 19\\
  \hline  
  \hline			
\end{tabular}
\end{table*}

\begin{figure}
\centering
\begin{minipage}{0.6\textwidth}
   \centering
  \includegraphics[width=\linewidth]{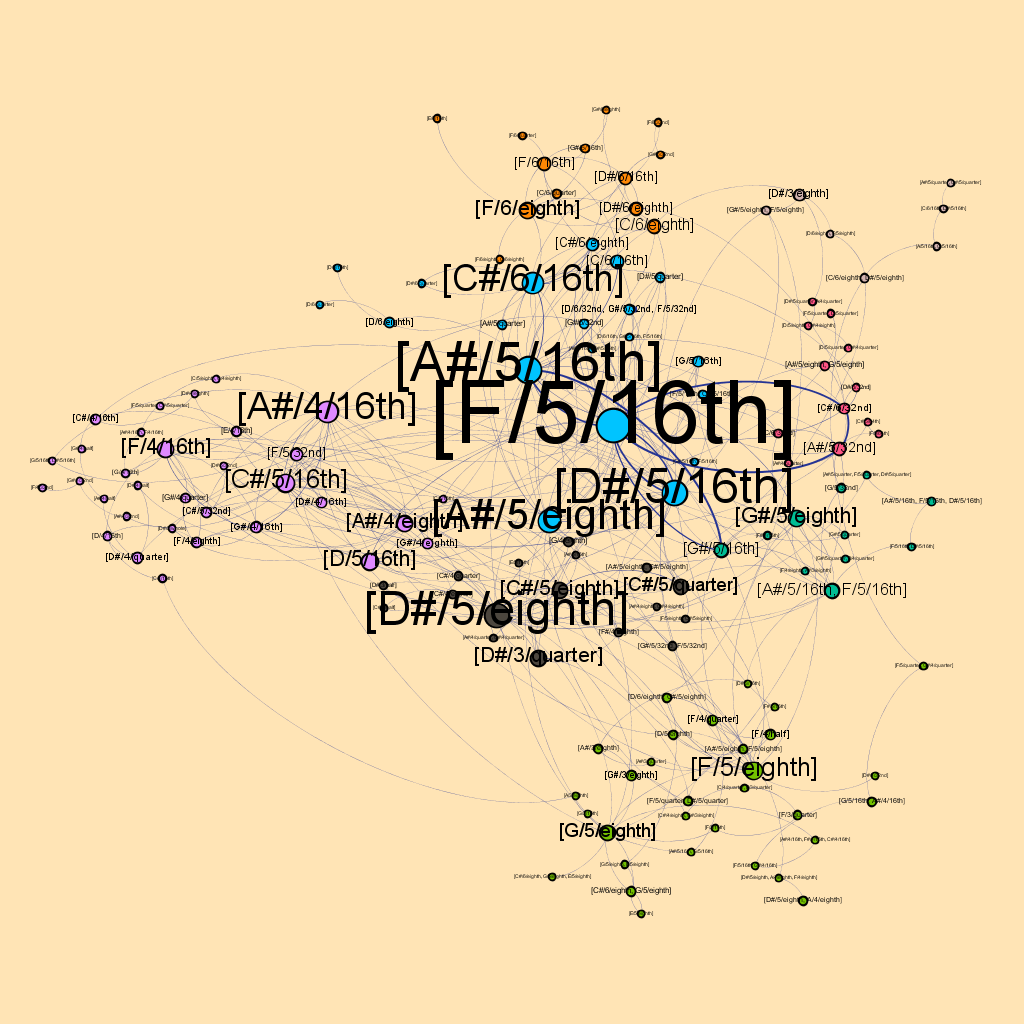}
\end{minipage}%
\begin{minipage}{0.4\textwidth}
   \centering
  \includegraphics[width=\linewidth]{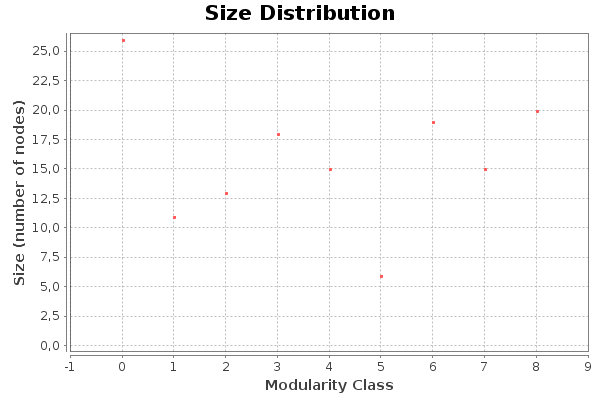}
\end{minipage}%
\caption{Modularity -- Jimi Hendrix; Red House. Rests have been removed from the track; nodes are colored according to their community.}
\label{fig:hendrix_modularity}
\end{figure}

\begin{figure}
\centering
\begin{minipage}{0.6\textwidth}
  \centering
  \includegraphics[width=\linewidth]{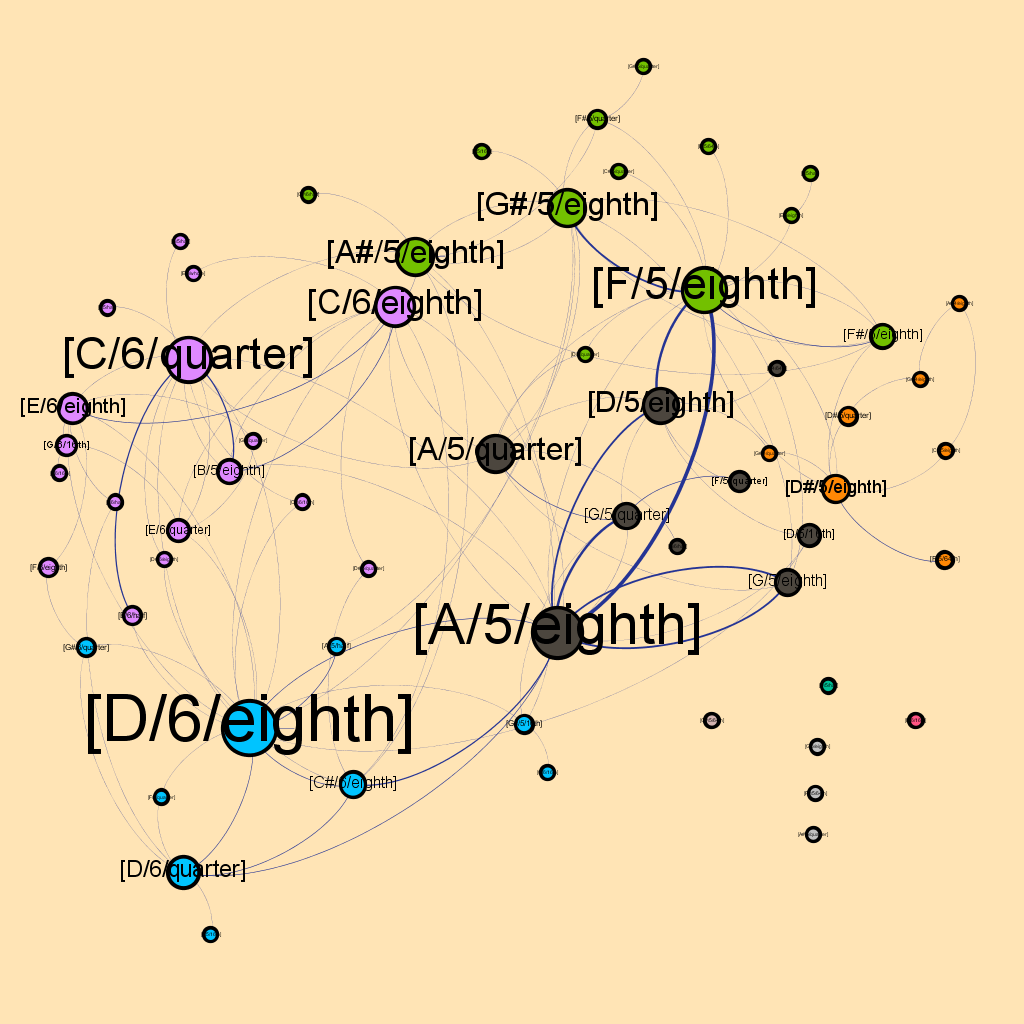}
\end{minipage}%
\begin{minipage}{0.4\textwidth}
    \centering
   \includegraphics[width=\linewidth]{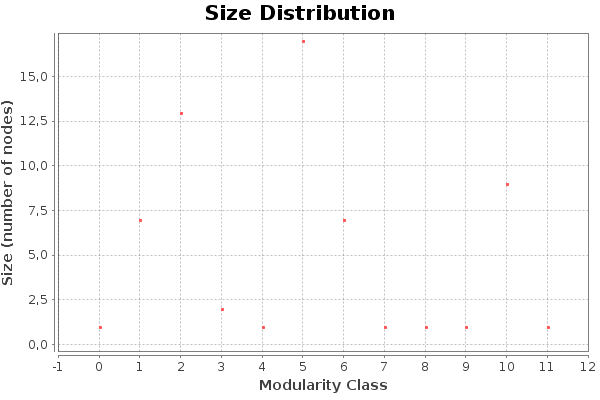}
\end{minipage}%
  \caption{Modularity -- Miles Davis; So What. Rests have been removed from the track; nodes are colored according to their community.}
  \label{fig:davis_modularity}
\end{figure}

\begin{figure}
\centering
\begin{minipage}{0.6\textwidth}
  \centering
  \includegraphics[width=\linewidth]{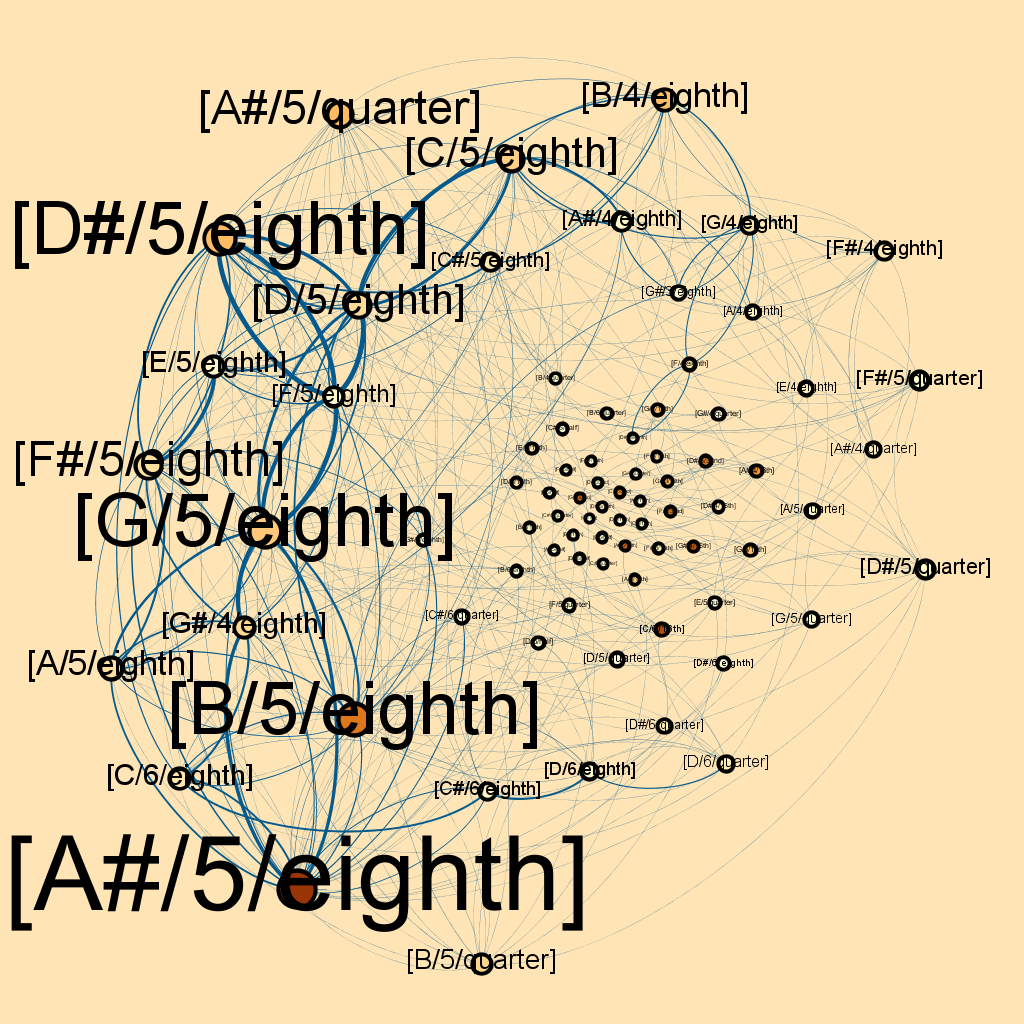}
\end{minipage}%
\begin{minipage}{0.4\textwidth}
    \centering
   \includegraphics[width=\linewidth]{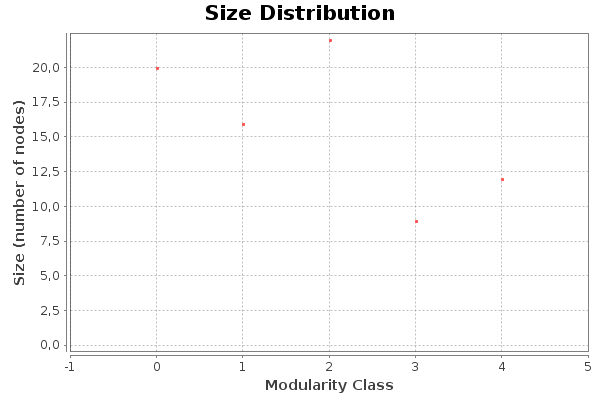}
\end{minipage}%
  \caption{Modularity -- John Coltrane; Giant Steps. Rests have been removed from the track; nodes are colored according to their community.}
  \label{fig:coltrane_modularity}
\end{figure}

\begin{figure}
\centering
 \begin{minipage}{0.6\textwidth}
    \centering
   \includegraphics[width=\linewidth]{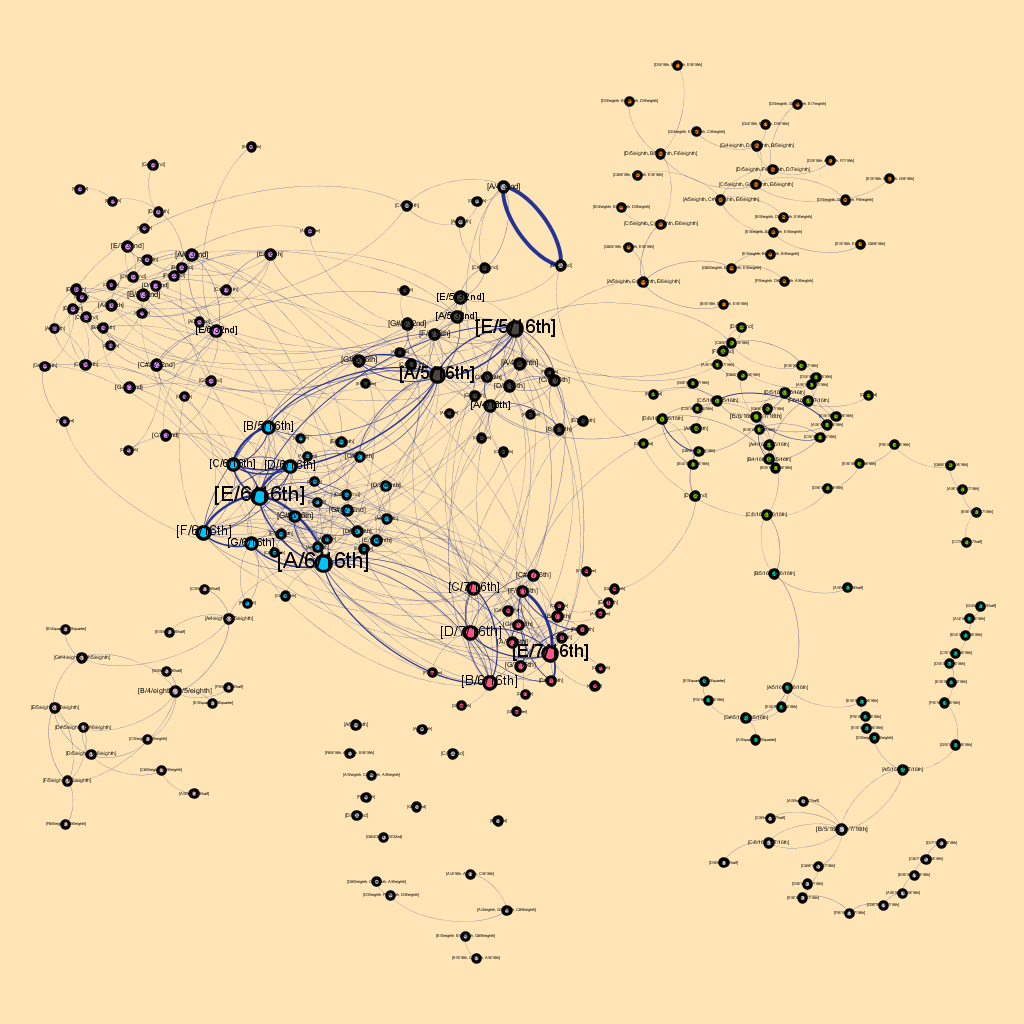}
 \end{minipage}%
 \begin{minipage}{0.4\textwidth}
    \centering
   \includegraphics[width=\linewidth]{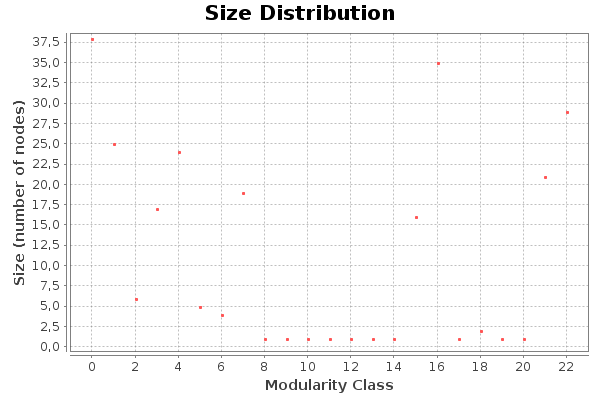}
 \end{minipage}%
  \caption{Modularity -- Niccolò Paganini; Caprice no.~24. Rests have been removed from the track; nodes are colored according to their community.}
  \label{fig:paganini_modularity}
\end{figure}

\begin{figure}
 \centering
 \begin{minipage}{0.6\textwidth}
    \centering
   \includegraphics[width=\linewidth]{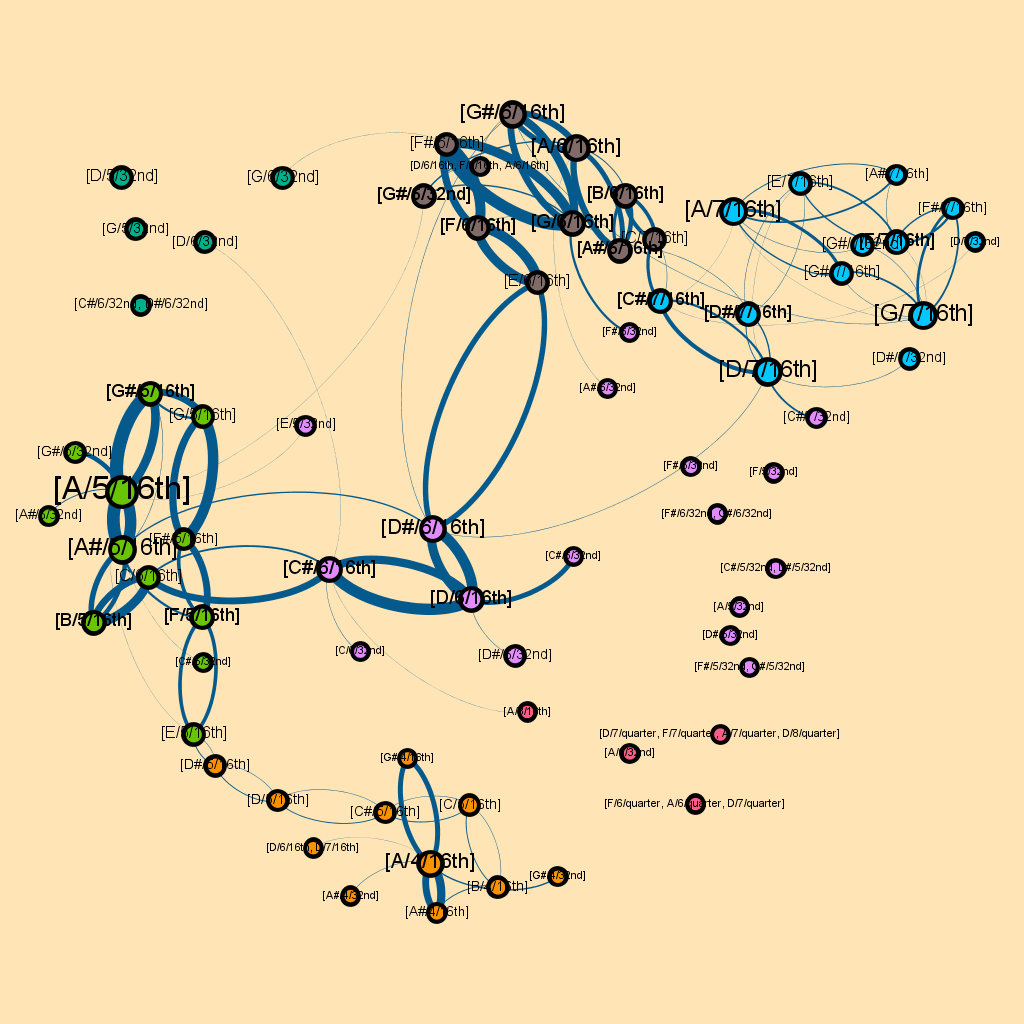}
 \end{minipage}%
 \begin{minipage}{0.4\textwidth}
    \centering
   \includegraphics[width=\linewidth]{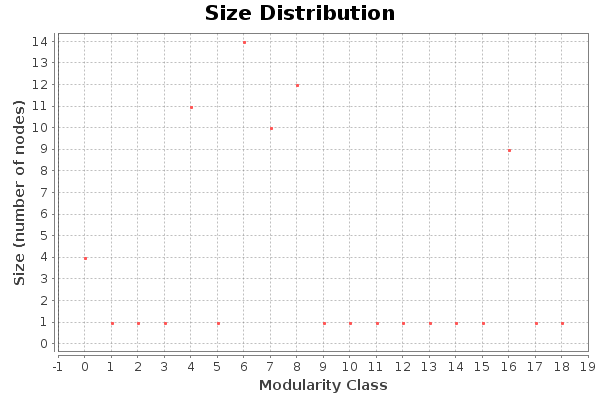}
 \end{minipage}%
  \caption{Modularity -- Nikolai Rimsky-Korsakov; Flight of the Bumblebee. Rests have been removed from the track; nodes are colored according to their community.}
  \label{fig:flight_modularity}
\end{figure}

This result can be appreciated also by looking at Figures \ref{fig:hendrix_modularity}--\ref{fig:flight_modularity}.
Each figure corresponds to a track. On the left side, a pictorial representation of the network is reported, where rests have been removed and nodes have been arranged in the layout and colored according to their community (thus, while networks are the same of the previous ones without rests, the layout and their resulting appearance are different). On the right side, the chart reports the size distribution of each community in the network, i.e.,~number of nodes composing each community.

Figure \ref{fig:hendrix_modularity} shows the network associated to the solo of J.~Hendrix in the ``Red House'' track, with nodes organized in communities. 
The network has several communities composed of a single node. It is possible to notice that even if some communities appear to be densely connected, an important amount of inter-community links exists. This is confirmed by the limited modularity value reported in Table \ref{tab:modularity} for that track.
We have a similar outcome for the solo played by M.~Davis in ``So What'' (Figure \ref{fig:davis_modularity}).
In this case, there are several communities that refer to isolated nodes, meaning that the musician played the corresponding notes alone, placing rests before and after that note.


A ``non-community structure'' is even more evident in ``Giant Steps'' (Fi\-gu\-re \ref{fig:coltrane_modularity}). Few communities are identified and the modularity value is lower than other considered networks (see Table \ref{tab:modularity}). In this case, the inter-community links are quite high; thus, the modularity algorithm fails in identifying sub-communities, which are indeed not present in the network.

A different outcome is obtained for the classical tracks ``Caprice no.~24'' (Figure \ref{fig:paganini_modularity}) and ``Flight of the Bumblebee'' (Figure \ref{fig:flight_modularity}). In this case, the mo\-du\-larity values are higher than others, as confirmed in Table \ref{tab:modularity}. Moreover, the network figures reveal a community-based structure, with few links connecting different communities. Also in this case, there are several isolated notes, or groups of notes, meaning that these are confined by two rests preceding and following them.


In conclusion, the study on modularity reveals that different tracks have different outcomes in terms of partitioning into communities. The considered classical tracks do have a community structure, while this aspect is not that evident in other (jazz) tracks. 
An interesting question is whether this feature depends on the music genre. The aim of this study was to demonstrate that a number of musical features can be obtained from a network-based analysis. The low number of considered tracks does not allow answering this question. Anyway, this analysis methodology can be employed to make detailed studies on the musical characteristics of different songs or melodies in general.

\section{Experimental Evaluation}\label{sec:eval}

This section shows aggregate results of some main metrics of interest related to the network-based analysis of musical tracks.
The employed data set is a bunch of a large set of musical solos of different guitar players \cite{Ferretti2017271}. 
Scores were retrieved from Web sites (e.g.~A-Z Guitar Tabs) in Guitar Pro or Power Tab formats. Thanks to the PyGuitarPro python library \cite{pyguitar}, and an own made (Python and Java) software, solo guitar parts were isolated, exported to a musicXML format \cite{musicXML}, and then exploited to create a network representation of the melodic line. 

\subsection{Density and Cumulative Distribution of Metrics of Interest}

We focus here on the density and cumulative distributions of the previously considered network metrics.

\begin{figure}
\centering
\begin{minipage}{0.5\textwidth}
   \centering
  \includegraphics[width=\linewidth]{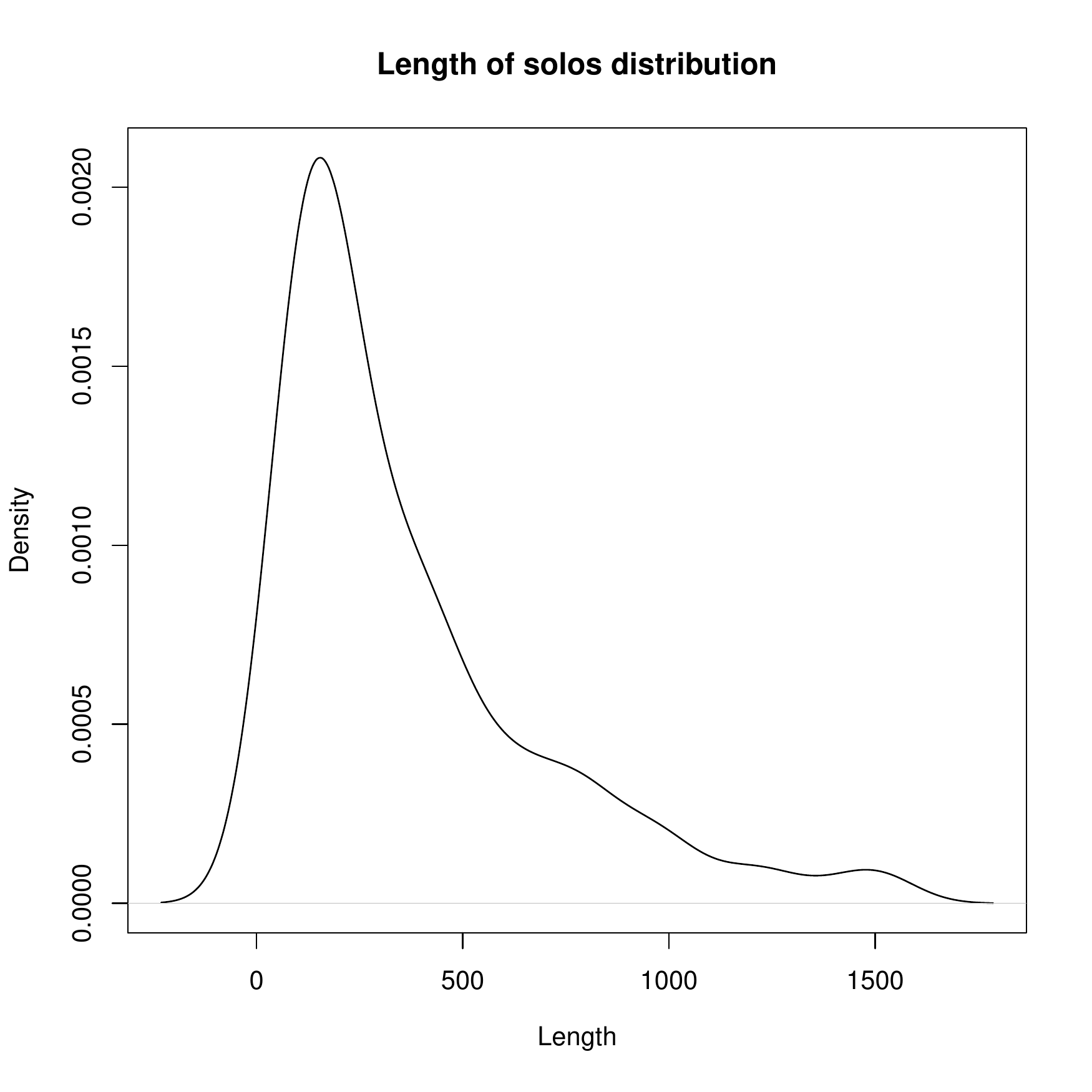}
\end{minipage}%
\begin{minipage}{0.5\textwidth}
   \centering
  \includegraphics[width=\linewidth]{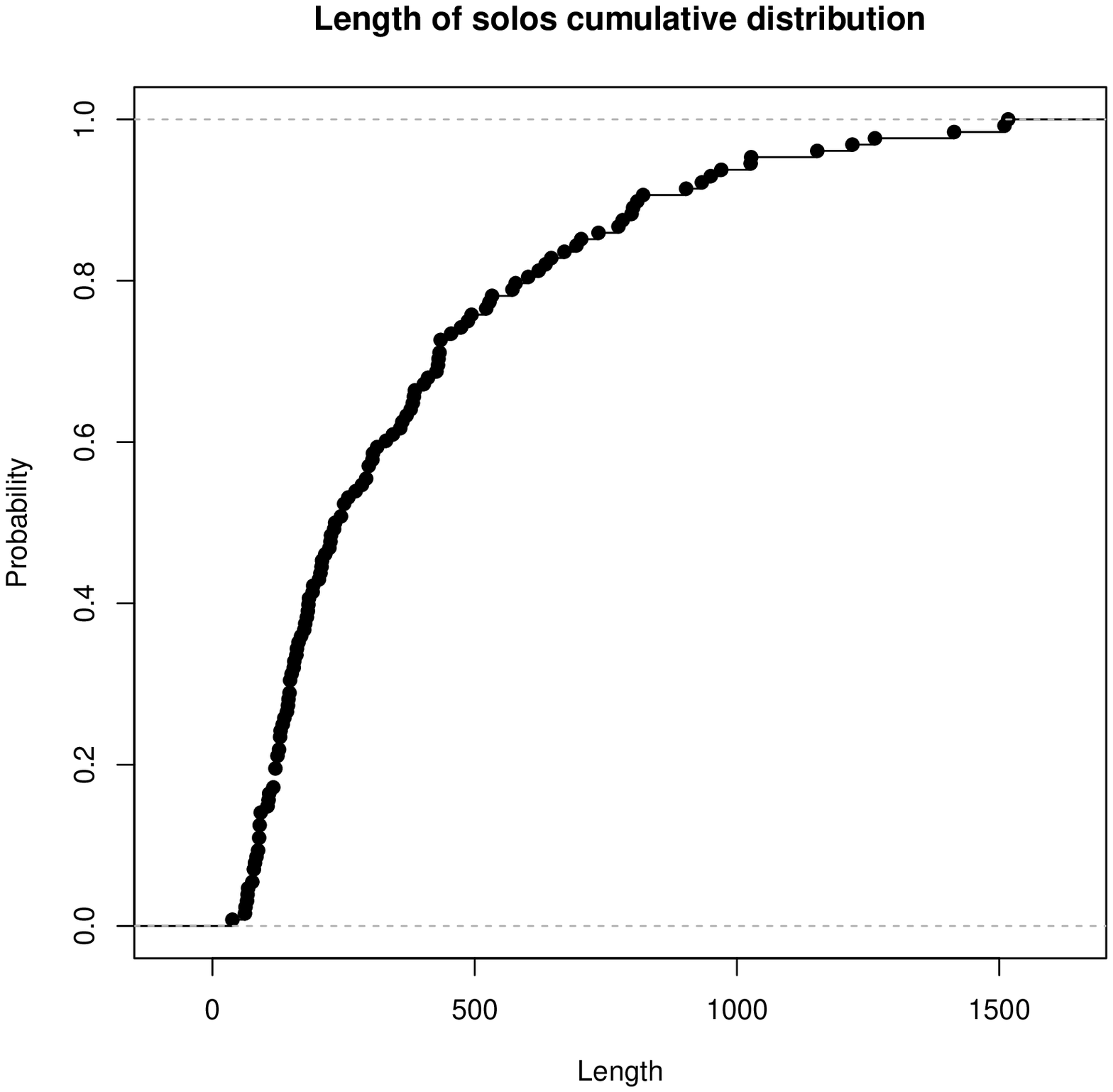}
\end{minipage}%
\caption{Length of solo distribution}
\label{fig:length}
\end{figure}

Figure \ref{fig:length} shows the density distribution and the cumulative distribution of the length of the considered solos. As mentioned, this metrics corresponds to the number of notes composing each solo. The density distribution (left-most chart) has a peak, with a heavy tail on the right part of the chart. This demonstrates that a variable amount of notes is played during a melodic line such as a musical solo.
The same outcome is confirmed by the cumulative distribution (right-most chart), showing a slow rise above $0.8$.

\begin{figure}
\centering
\begin{minipage}{0.5\textwidth}
   \centering
  \includegraphics[width=\linewidth]{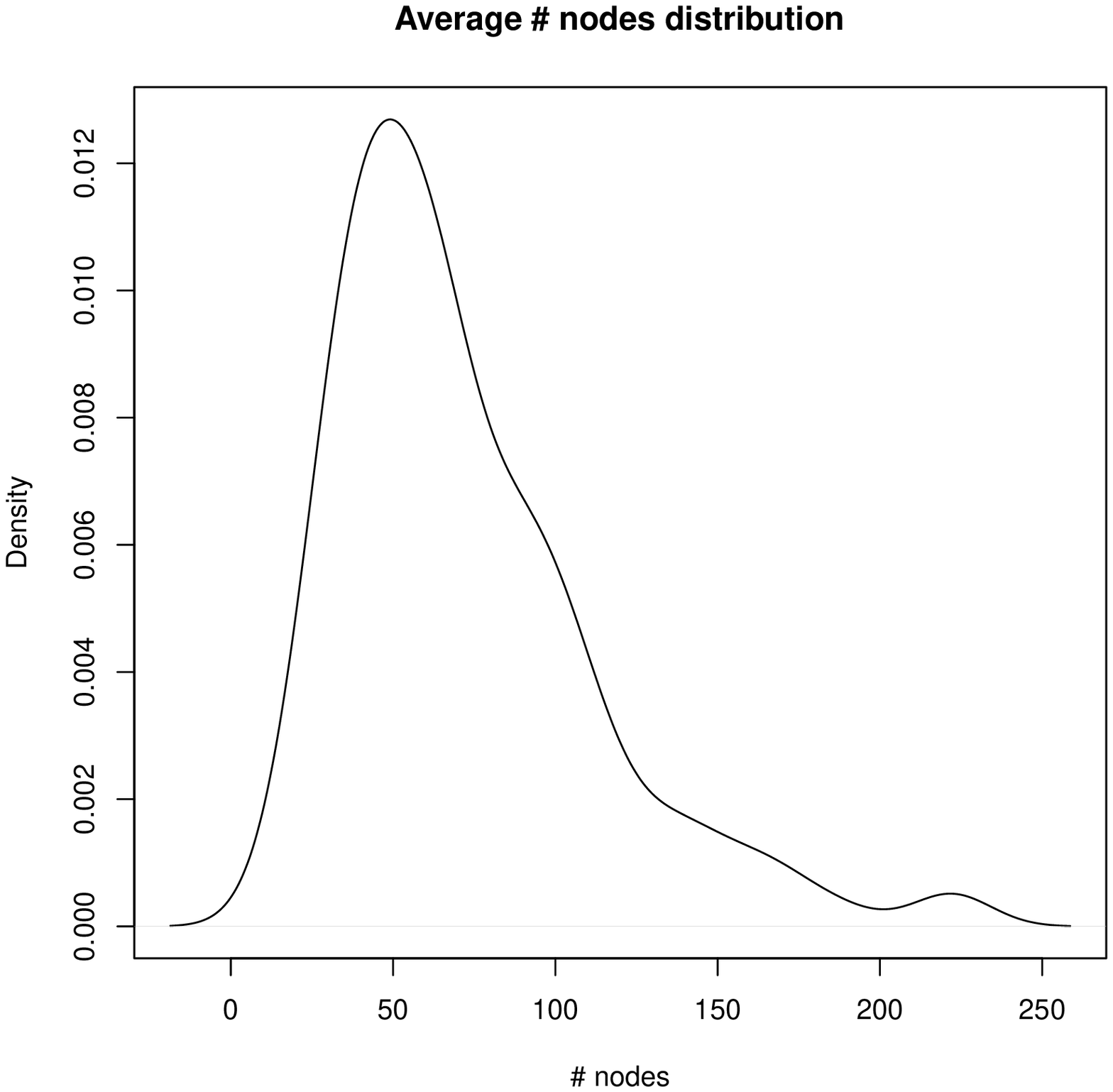}
\end{minipage}%
\begin{minipage}{0.5\textwidth}
   \centering
  \includegraphics[width=\linewidth]{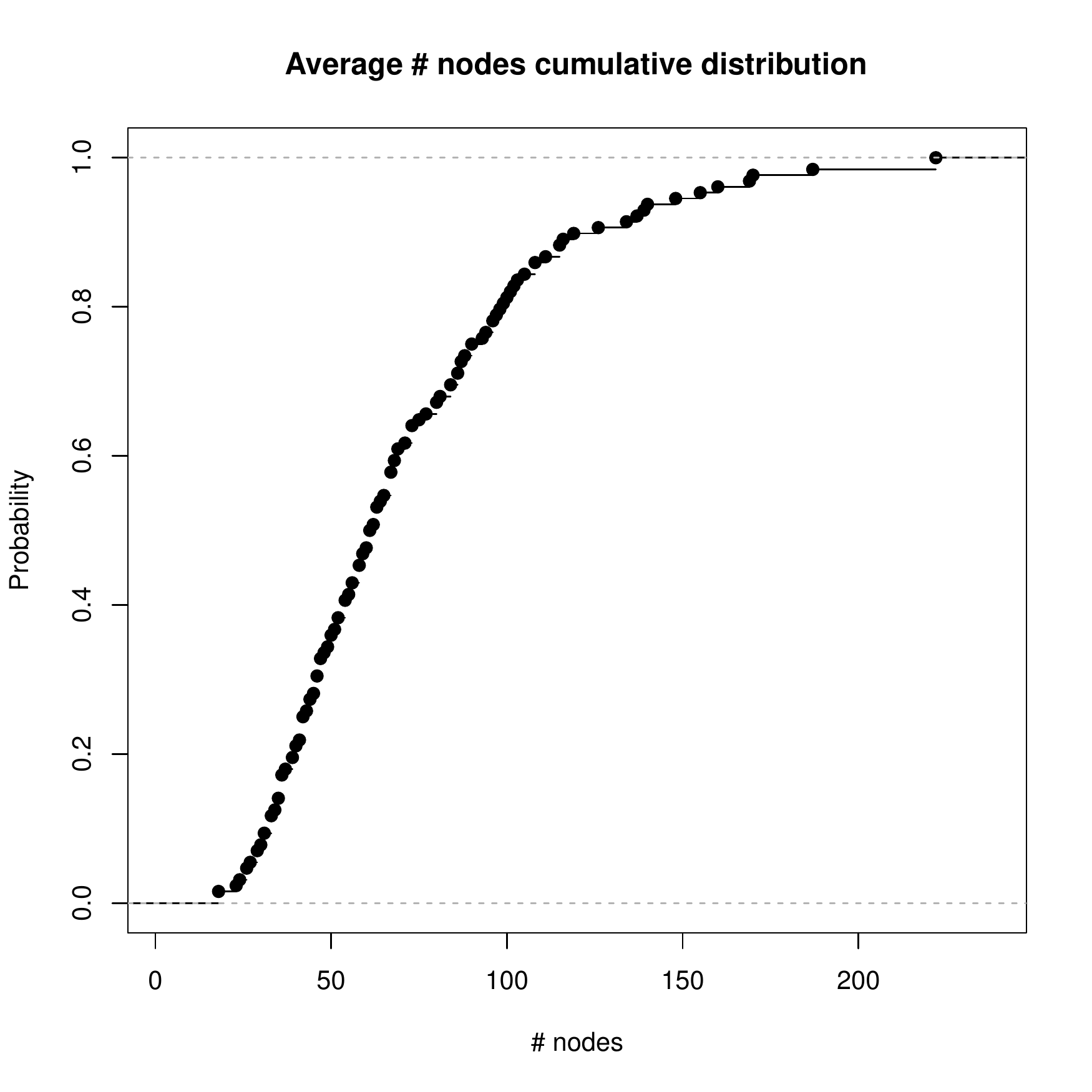}
\end{minipage}%
\caption{Average number of nodes distribution}
\label{fig:nodes}
\end{figure}

Figure \ref{fig:nodes} shows the distribution of the amount of nodes. It is possible to notice a similar trend to the previous chart, even if values are lower. Indeed, the two considered metrics are somehow correlated, since in a melodic line the same note/node is usually played more that once.

\begin{figure}
\centering
\begin{minipage}{0.5\textwidth}
   \centering
  \includegraphics[width=\linewidth]{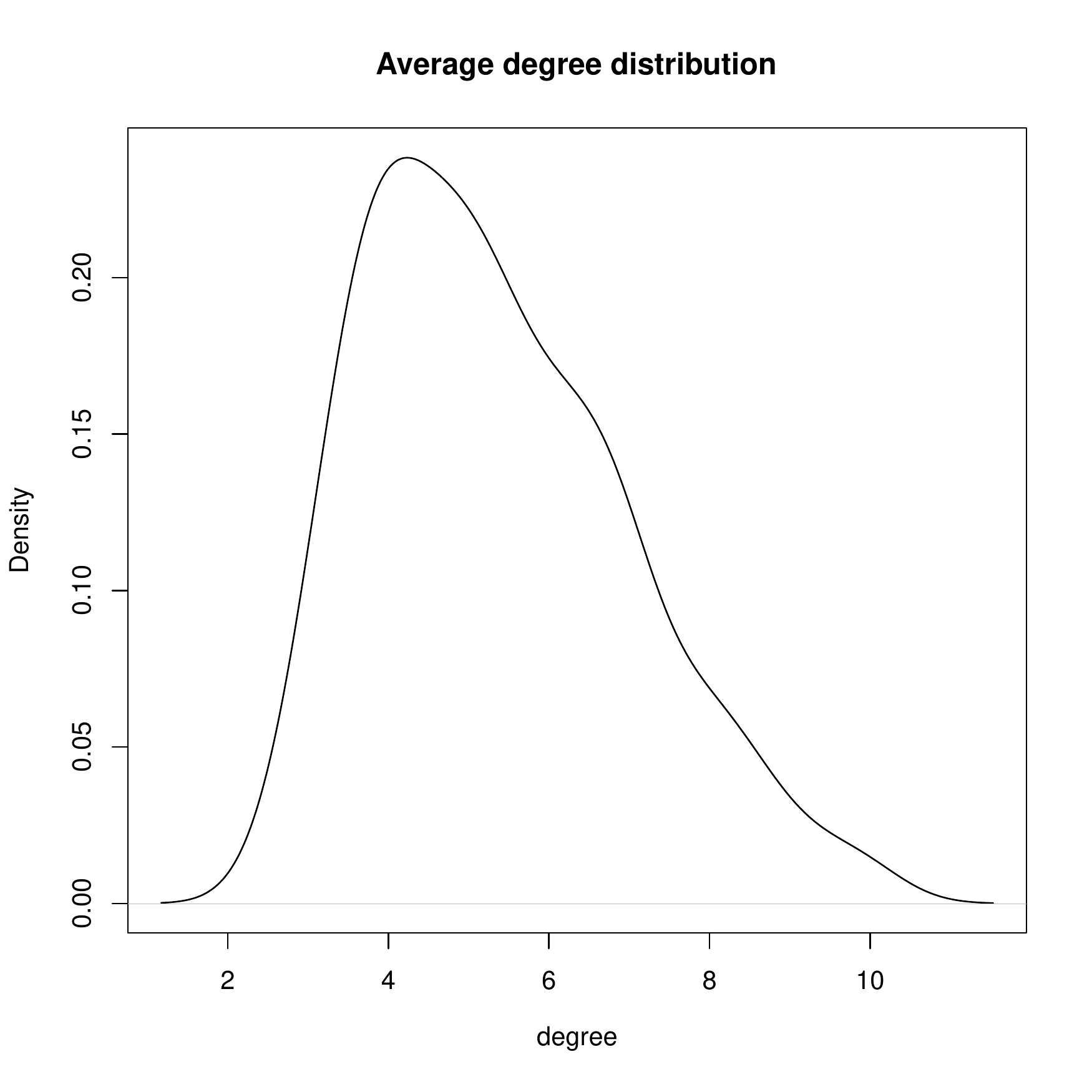}
\end{minipage}%
\begin{minipage}{0.5\textwidth}
   \centering
  \includegraphics[width=\linewidth]{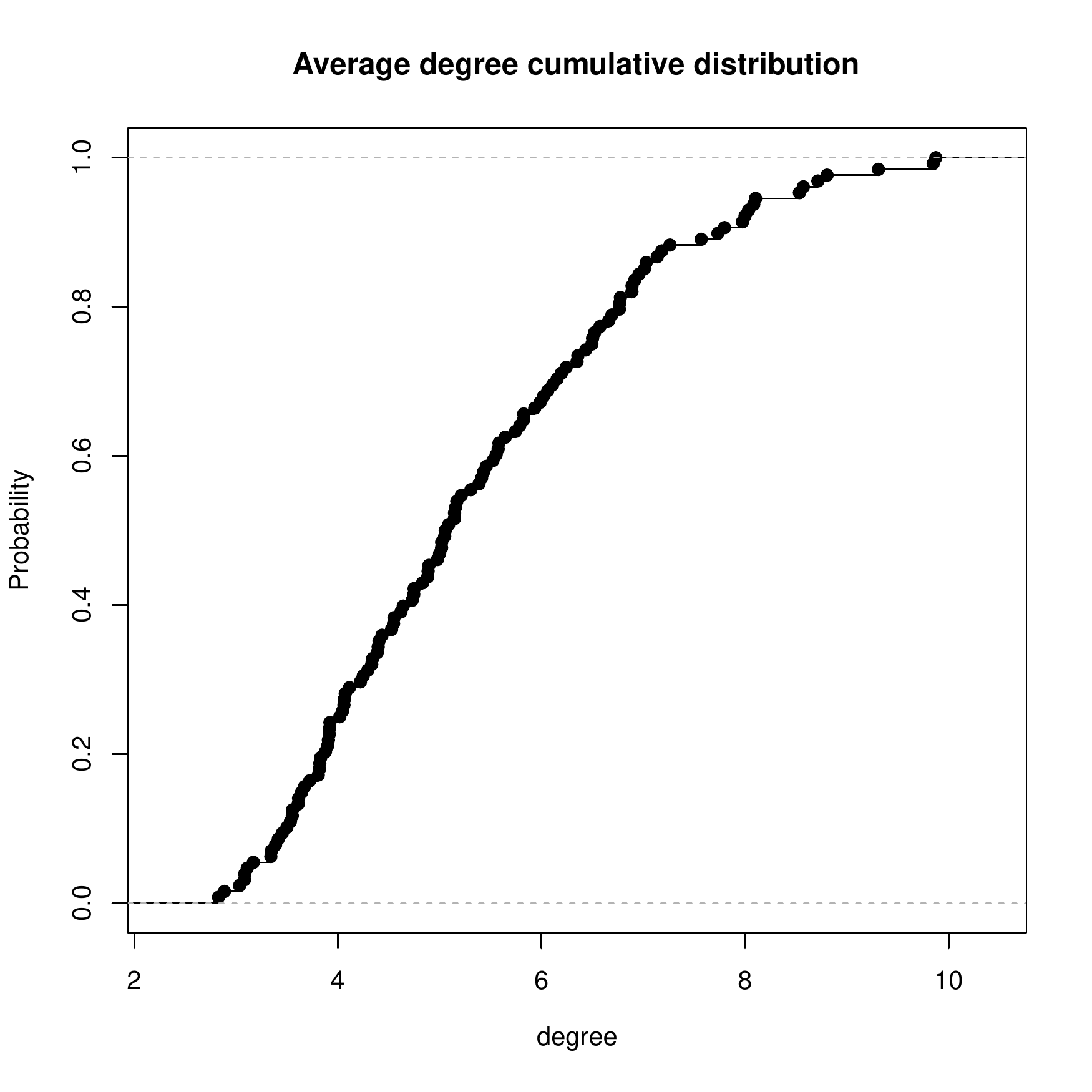}
\end{minipage}%
\caption{Average degree distribution}
\label{fig:degree}
\end{figure}

Figure \ref{fig:degree} shows the distribution of the average degree of networks obtained from the considered melodic lines. 
The average degree distribution has a peak value around $4$ and the average degrees lie in general between $2$ and $11$. 
A very similar trend is evident when looking at in-degrees degrees (not reported here).
This outcome demonstrates that in a melody notes are played before/after multiple other ones in a melody, confirming that a peculiar (and well known) aspect of the construction of a melody is on how to combine notes.

\begin{figure}
\centering
\begin{minipage}{0.5\textwidth}
   \centering
  \includegraphics[width=\linewidth]{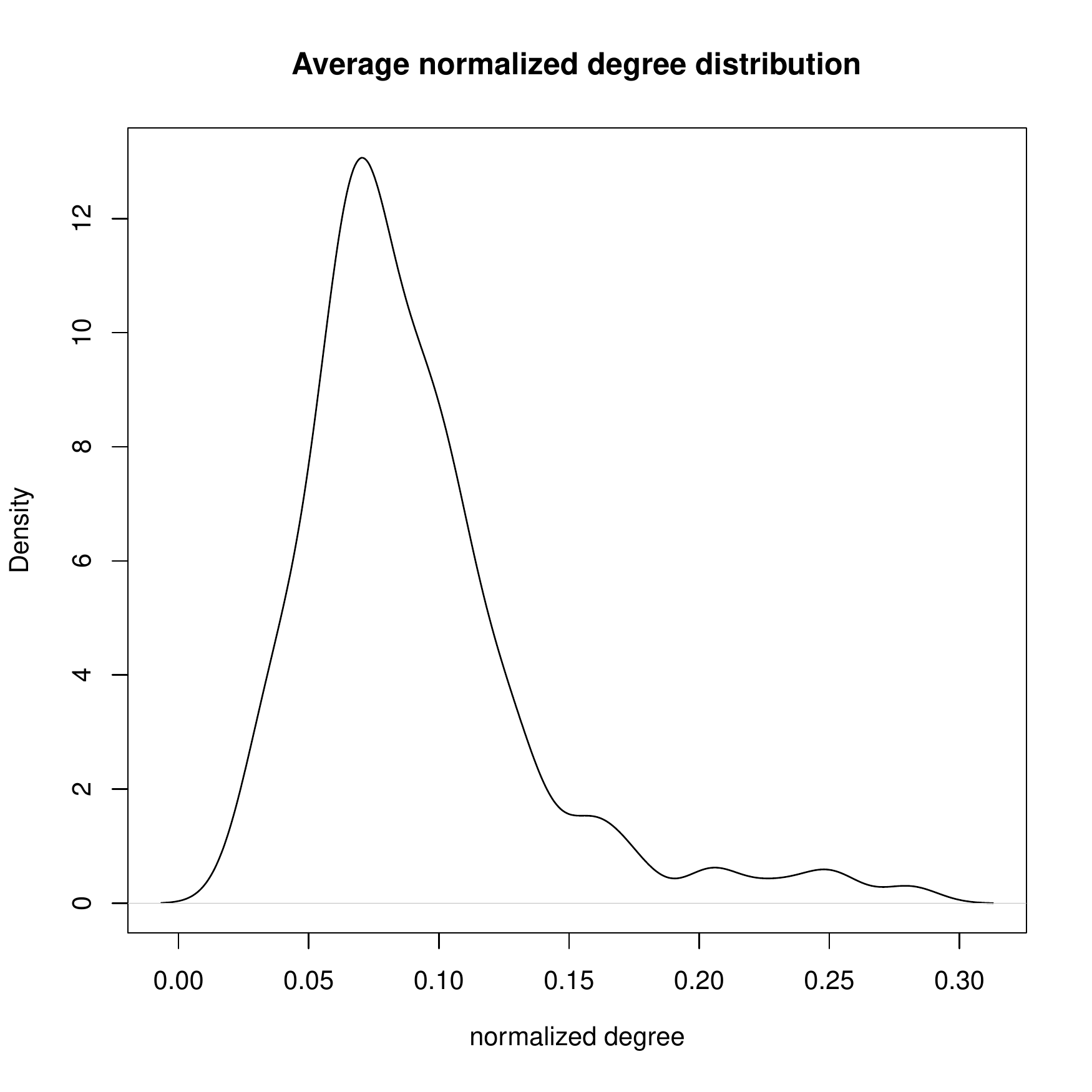}
\end{minipage}%
\begin{minipage}{0.5\textwidth}
   \centering
  \includegraphics[width=\linewidth]{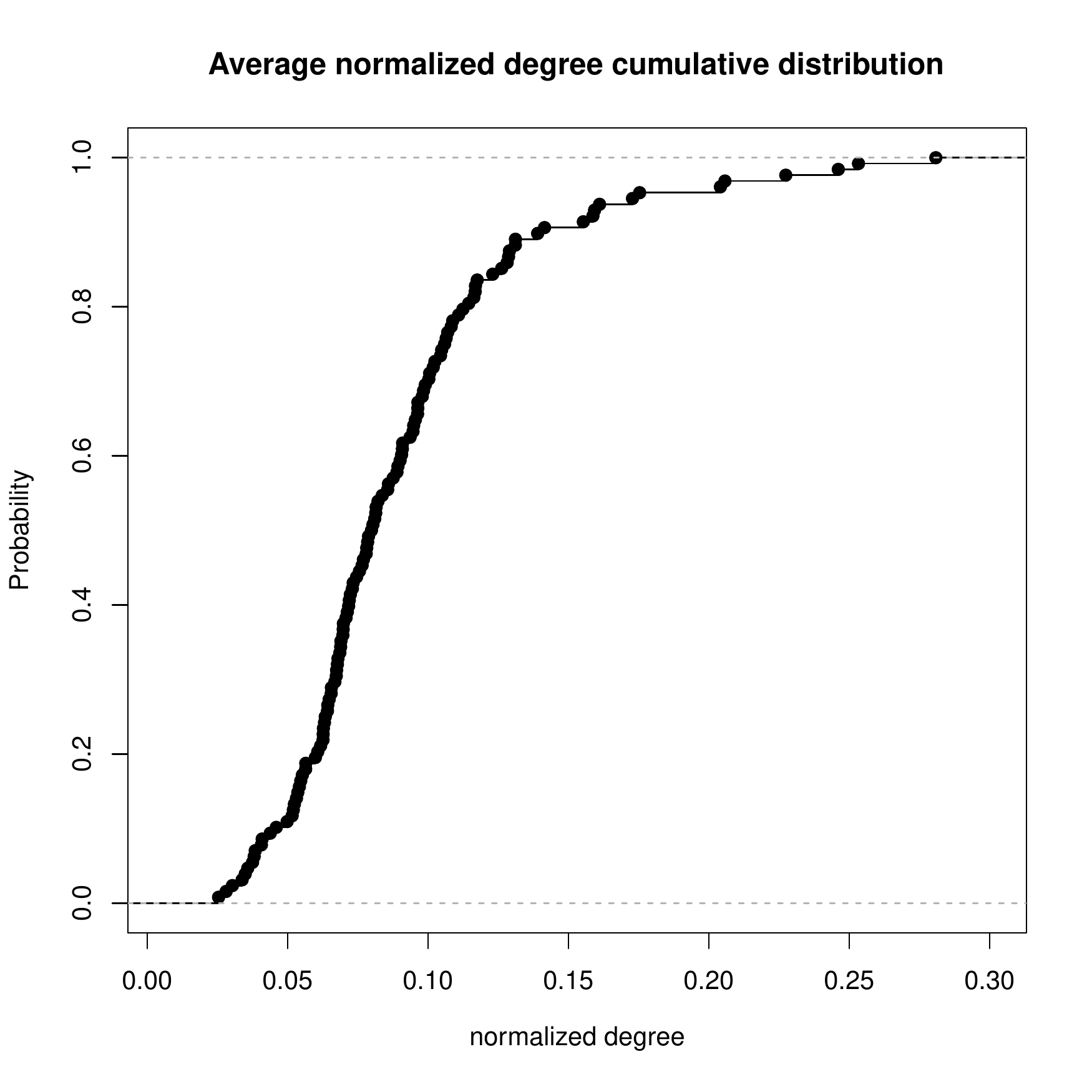}
\end{minipage}%
\caption{Average normalized degree distribution}
\label{fig:normalized_degree}
\end{figure}

Figure \ref{fig:normalized_degree} shows the density and cumulative distributions of normalized degrees. In this case, the bell curve of the density distribution is more narrow with respect to the average degree; a large tail is present for higher values.

\begin{figure}
\centering
\begin{minipage}{0.5\textwidth}
   \centering
  \includegraphics[width=\linewidth]{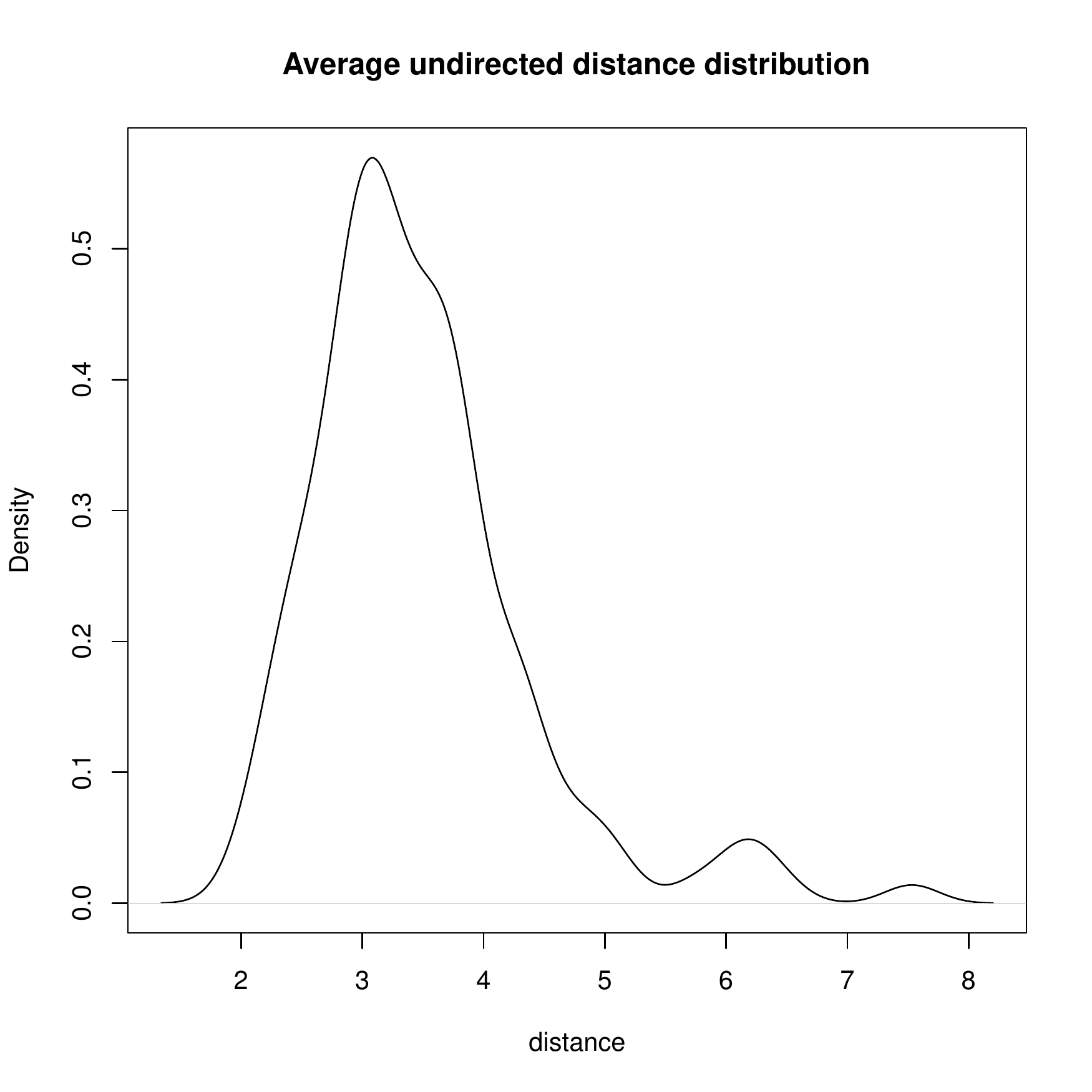}
\end{minipage}%
\begin{minipage}{0.5\textwidth}
   \centering
  \includegraphics[width=\linewidth]{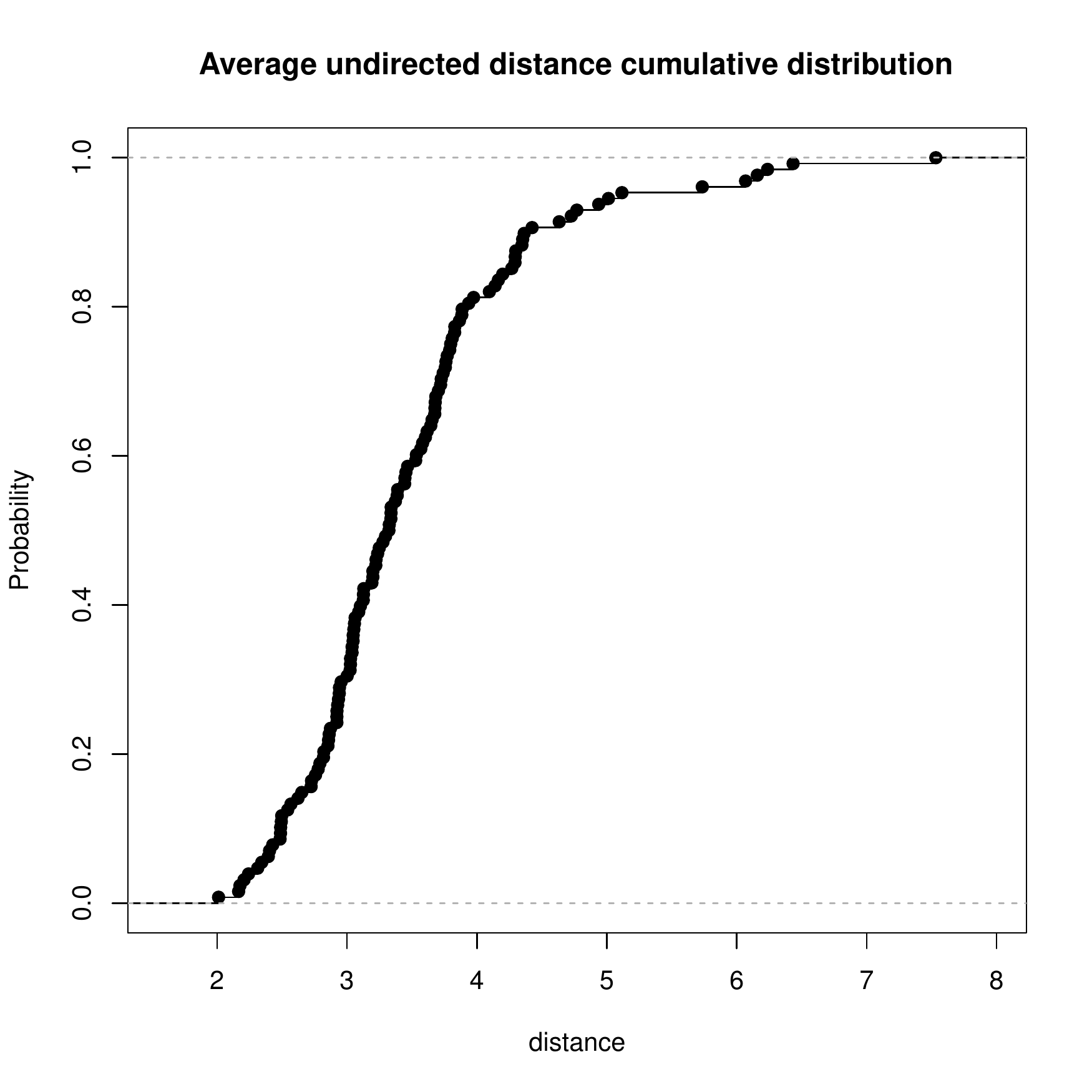}
\end{minipage}%
\caption{Undirected distance distribution}
\label{fig:distance}
\end{figure}

Figure \ref{fig:distance} shows the density and cumulative distributions of the average undirected distances of nodes in the considered nets.
Again, density distribution is a bell curve with a larger tail for higher distance values. The cumulative distribution confirms that musical networks have a low average path length, in general. For instance, the $60\%$ of the considered networks has an average distance among nodes which is lower than $4$. This means that, in the considered melodic lines played/generated by the the musicians/composers, the majority of exploited notes are not far away from others. 

\begin{figure}
\centering
\begin{minipage}{0.5\textwidth}
   \centering
  \includegraphics[width=\linewidth]{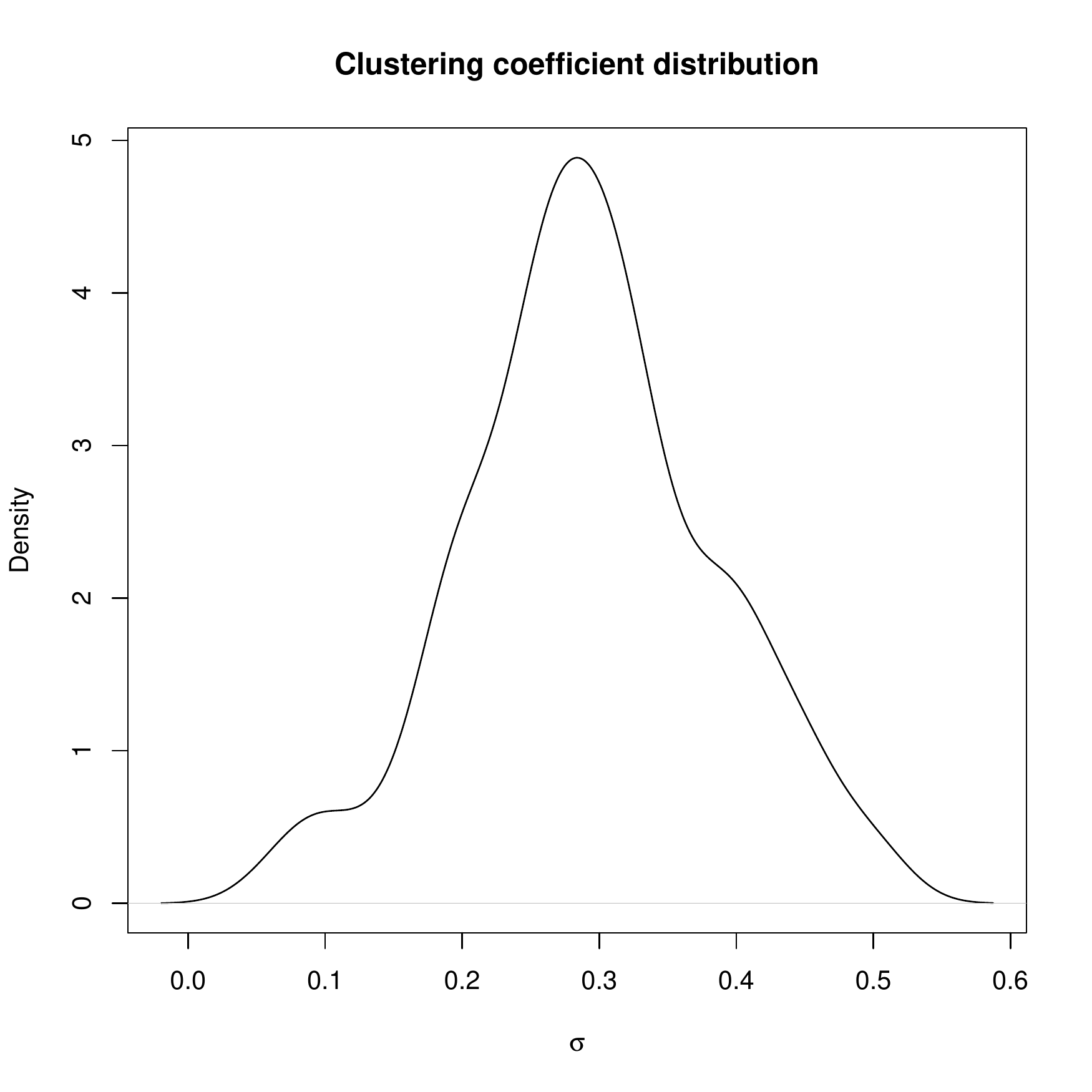}
\end{minipage}%
\begin{minipage}{0.5\textwidth}
   \centering
  \includegraphics[width=\linewidth]{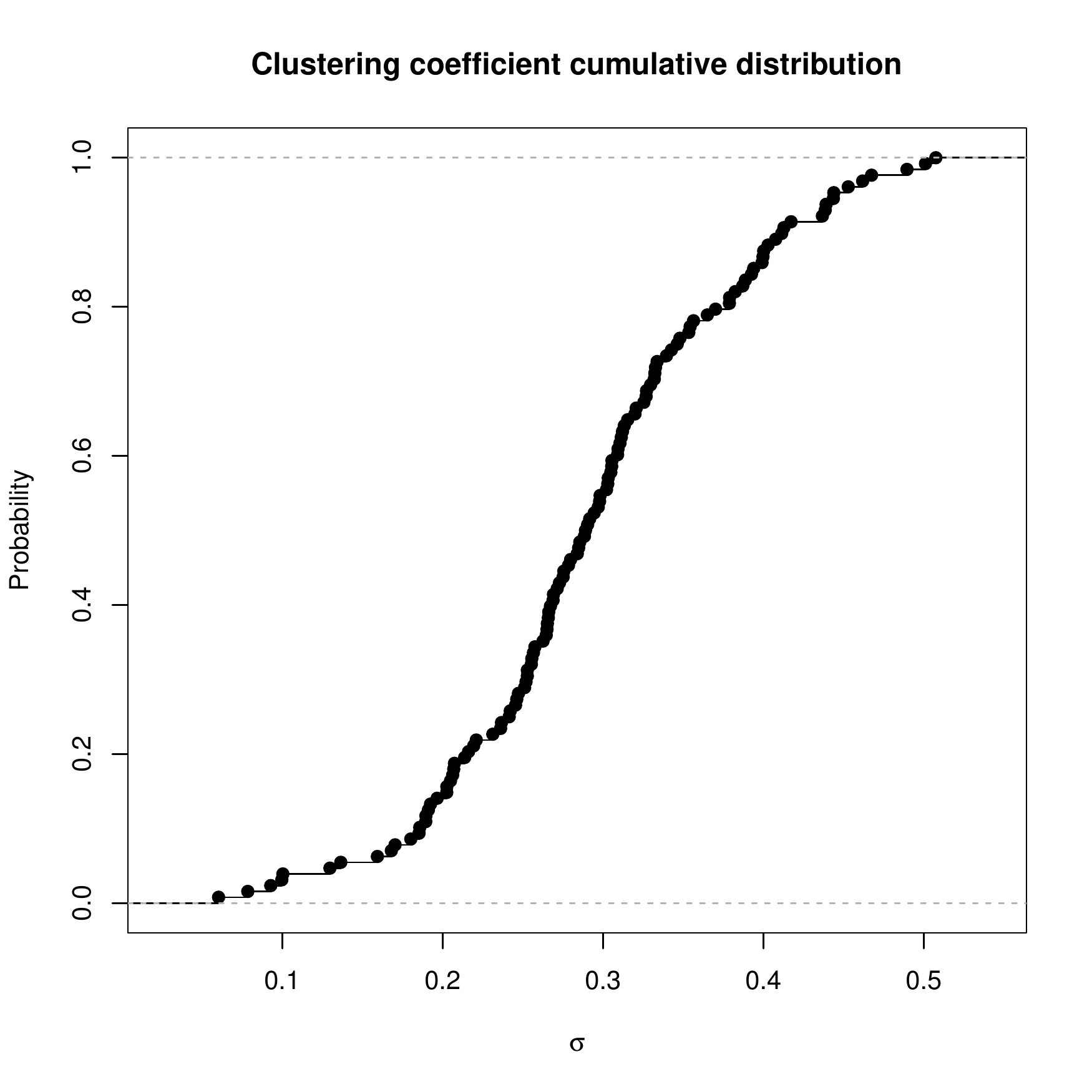}
\end{minipage}%
\caption{Clustering coefficient distribution}
\label{fig:cc}
\end{figure}

Figure \ref{fig:cc} is concerned with the distribution of the clustering coefficients.
We already mentioned that the clustering coefficient states how much notes are clustered, i.e., how much the performer plays notes in an interchangeable order. This value is comprised between $0$ and $1$, where $0$ means that no triangles are present in the network, i.e.,~notes are played in a specific sequence, and $1$ means that all triplets of nodes form a triangle, i.e., all triplets of notes of the melody are played in all possible orders. 
In a network representing a melody, it is unlikely to have a fully connected network with an overall clustering coefficient equal to $1$. Thus, a peak around $0.3$ in the density distribution of the clustering coefficient confirms that melodies are well clustered. This is in accordance to the specific network/melody exemplars analyzed in Section \ref{sec:ex}.

\begin{figure}
\centering
\begin{minipage}{0.5\textwidth}
   \centering
  \includegraphics[width=\linewidth]{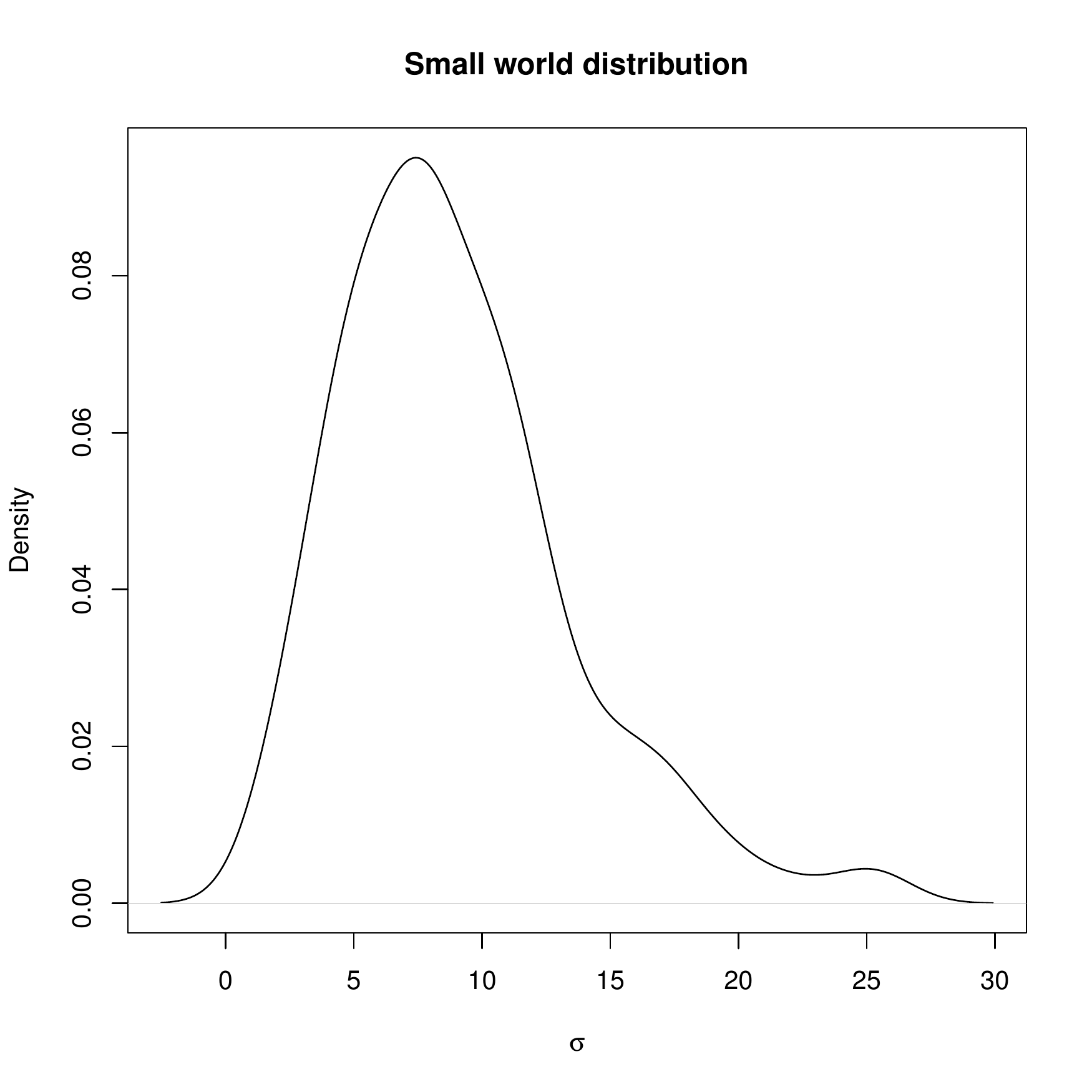}
\end{minipage}%
\begin{minipage}{0.5\textwidth}
   \centering
  \includegraphics[width=\linewidth]{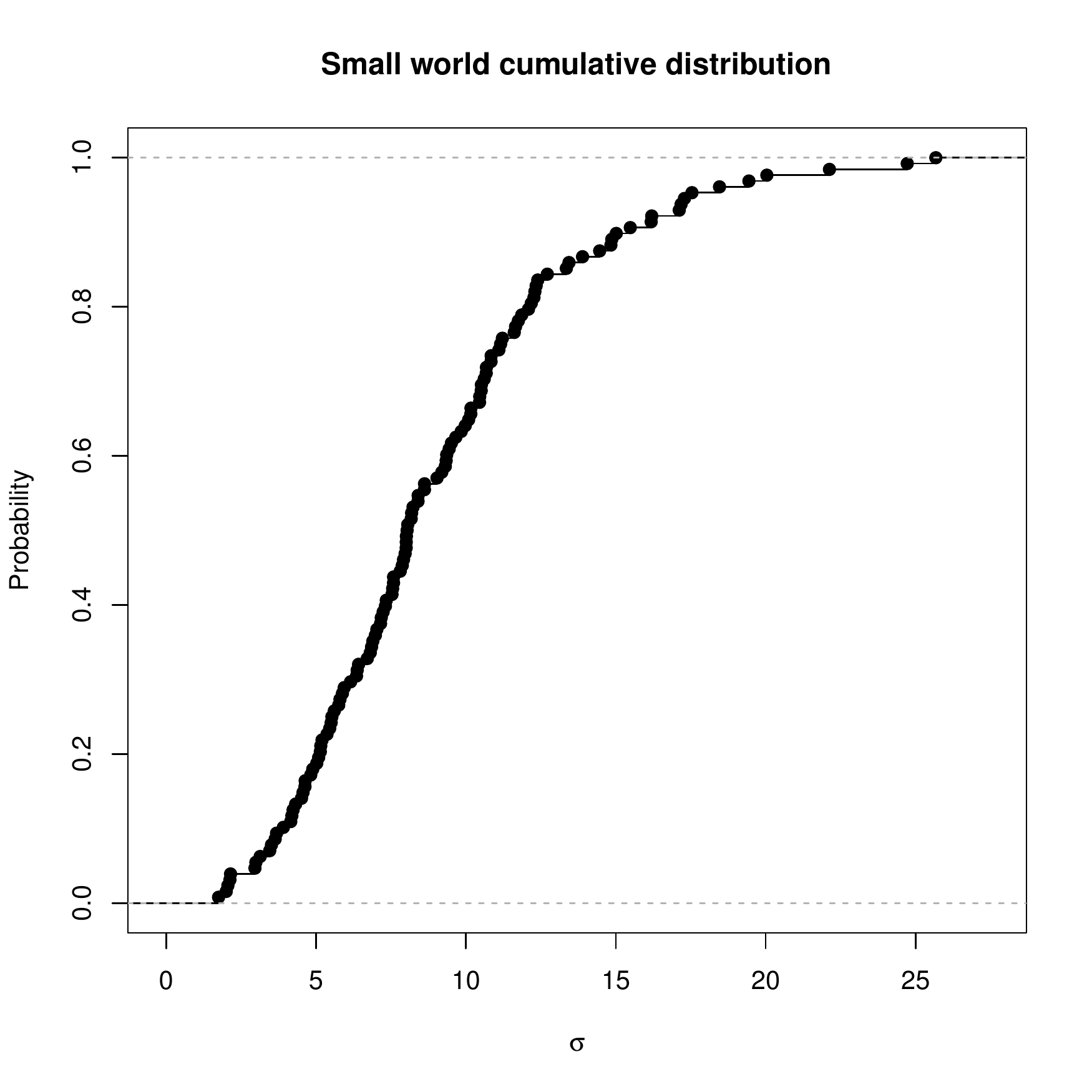}
\end{minipage}%
\caption{Small World $\sigma$ value distribution}
\label{fig:sw}
\end{figure}

Finally, Figure \ref{fig:sw} reports the density and cumulative distributions of the small world $\sigma$ values ($\sigma > 1$ implies that the network is a small world). The charts confirm that almost all melodies/solos exhibit the small world property, meaning that it is a general trend to exploit locality in the networks (i.e., groups of notes are played in an interchangeable order in the melody), but there are links (sequences of two notes) connecting more distant groups of notes in the net. Such connecting links might include the so called ``passing notes'', i.e.~melodic embellishments that occur between two stable notes (typically chord tones), creating stepwise motion. 

All these results demonstrate that these networks have common properties in general, however with some differences that can be studied and used to characterize the peculiarity of a song, a musician or composer. For instance, clustering techniques, or alternative machine learning approaches, might be employed to identify those musical tracks (or musicians) that have similar musical traits.

\subsection{Comparison with Real Networks}

\begin{table*}[th]
\centering
\caption{Comparison among some musical networks and some well known real networks, of similar size.}
\label{tab:comp_real_nets}
\scriptsize
\begin{tabular}{|| l || c | c | c ||}
  \hline  
  \hline			
  network &                                       \#nodes  & avg deg  & $\sigma$\\
  \hline  
  \hline			
   J.~Hendrix -- Red House                       & 148   & 6.19     & 17.8\\
   \hline
   J.~Hendrix -- Voodoo Child (Slight Return) -- Woodstock & 222	& 5.98	& 32.7\\
   \hline			
   N.~Paganini -- Caprice no.~24                 & 257   & 5.87     & 29.9\\
  \hline  
  \hline  
   E.~Coli -- substrate graph \cite{Fell2000}			& 282	& 7.35	& 12.9\\
  \hline  
   Silwood Park food web \cite{MONTOYA2002405}			& 154	& 4.75	& 4.7\\
  \hline  
   C.~Elegans \cite{watts1998cds}				& 282	& 14	& 4.7\\
  \hline  
  \hline			
\end{tabular}
\end{table*}

Table \ref{tab:comp_real_nets} compares three musical networks (first rows of the table) with three real networks that have been widely analyzed in the complex networks literature (latest three rows of the table).
These specific six networks were selected due to their similar size in terms of number of nodes (see ``\#nodes'' column in the table).

The three musical networks are the two already considered tracks ``Red House'' and ``Caprice no.~24'', plus the network obtained from the first solo improvisation performed by J.~Hendrix, over the song ``Voodoo Child (Slight Return)'', during his famous performance at Woodstock in 1969.

The other considered three networks are very diverse in nature. 
The first one is the substrate graph of the \textit{Escherichia Coli}, where nodes represent chemical compounds, which are connected in the network when they occur (either as substrates or products) in the same chemical reaction \cite{Fell2000}.
The ``Silwood Park food web'' is a graph representation of the predatory interactions among species in the Silwood Park environment. Each node represents a species, and a directed link is drawn from $i$ to $j$ when species $j$ preys on species $i$ \cite{MONTOYA2002405}.
Finally, the neural network of the nematode worm \textit{Caenorhabditis elegans} is an important example of a completely
mapped neural network. Nodes are neurons, and a link joins two neurons if they are connected by either
a synapse or a gap junction \cite{watts1998cds}.

While a detailed comparison is out of the scope of this paper, the rationale here is to emphasize that very diverse networks may have similar mathematical properties. In fact, the average degree for these networks is comparable, in the range of approximately $4.5$ up to $7.5$, with the exception of the ``C.~Elegans'' net, which has a higher value than others, i.e., $14$. However, in the previous charts it has been shown guitar solos in the considered dataset have an average degree equal or higher than $10$. 
Moreover, the table shows that all these different networks are small worlds; in fact, the $\sigma$ value is well above $1$ in all cases, thus demonstrating this common property for all of them.

This result shows that musical networks have structural properties similar to other complex networks, which model very diverse physical or biologic phenomena. This is a common trend in the graph modeling of real systems, which suggests that further reasoning, typical of complex network theory, can be proficiently applied to music as well. 

\section{Conclusions}\label{sec:conc}

In this paper we discussed an approach to model melodies as networks. According to the model, notes of a melody are treated as network nodes, and links are added between two nodes when the related notes are played in succession.
The pictorial representation of the network gives an idea on the network structure and its related track. However, it is possible to extract several metrics that provide interesting insights on the melody and its musical characteristics.
One can thus look at the length of the melody, node degrees and their distribution, average distance, betweenness, network density, clustering coefficient, modularity and so on. We can also understand which are the main notes in the network (melody) structure.

In general, we saw that melodies and solos, which are classified by music experts as very sophisticated ones, do have a corresponding complex network structure.
This is true, as an example, for the solo by J.~Coltrane in ``Giant Steps''.
Results suggest that most melodic networks are small worlds and exhibit a scale-free network structure.

Rests play an important role in music, regardless of the music genre or instrument being used (this is a confirmation of a straightforward, well known musical claim). Not surprisingly, rests play the role of hubs in many networks and may have high betweenness values.

Intricate melodies, typical of modern music, have a wide node degree distribution, with high median and average degrees, a high clustering coefficient, with some edge weights significantly higher than others. Conversely, the considered classical tracks, while recognized as very difficult from an execution point of view, have a more linear structure. In this case, the amount of hubs is lower, while certain edges have a high weight. 

The two considered classical tracks have a high modularity value. Without rests, the networks have several isolated nodes and a clear organization in communities. Such a feature is not that evident in other tracks.

The use of a mathematical modeling of a music track provides a general and compact way to analyze music.
This model can be of help in capturing the artistic traits of a musician, a music genre, the typical melodies obtained through a specific instrument, and so on.
The application scenarios are related to the classification and categorization of music. 
Similarity and clustering of musical tracks can be performed by comparing the mentioned metrics, as well as by employing more sophisticated network similarity algorithms \cite{citeulike:1319024,Ferretti2017271,ShahKZGF15}.

Moreover, the proposed framework can be exploited as a tool during the automatic generation of music. 
For instance, let consider the case of generating a melody whose style resembles that of a given musician.
The idea is that the networks' structure, together with the main parameters that describe these nets, can be employed to identify main motifs and peculiar characteristics of the musician. These can be utilized as inputs and used together with modern machine learning techniques to generate melodies, whose structure reflects the peculiar aspects of networks generated by the artist \cite{ref_nonsense}.
Such an approach might have interesting applications in music didactics, multimedia entertainment, and digital music generation.
All these applications are regarded as future work. In these scenarios, it will be also interesting to understand whether the representation of a music score as a network introduces additional benefits in these specific use cases, besides the possibility to characterize music and artists through a mathematical description.


\bibliographystyle{abbrv}

\end{document}